\documentclass[a4paper,11pt]{article}
\pdfoutput=1 % if your are submitting a pdflatex (i.e. if you have
\usepackage{jcappub} % for details on the use of the package, please
\usepackage{natbib}

\usepackage[T1]{fontenc} % if needed

\title{Searching for dark matter annihilation from individual halos: uncertainties, scatter and signal-to-noise ratios.}
\author[a,b]{Chiamaka Okoli,}
\author[b]{James E.~Taylor}
\author[a,b]{and Niayesh Afshordi}
\affiliation[a]{University of Waterloo,\\200 University Avenue West, Waterloo, ON N2L 3G1, Canada}
\affiliation[b]{Perimeter Institute for Theoretical Physics, \\ 31 Caroline Street North, Waterloo, ON N2L 2Y5, Canada}
\emailAdd{c2okoli@uwaterloo.ca}
\emailAdd{taylor@uwaterloo.ca}
\emailAdd{nafshordi@pitp.ca}

\abstract{Individual extragalactic dark matter halos, such as those associated with nearby galaxies and galaxy clusters, are promising targets for searches for gamma rays from dark matter annihilation. We review the predictions for the annihilation flux from individual halos, focusing on the effect of current uncertainties in the concentration-mass relation and the contribution from halo substructure, and also estimating the intrinsic halo-to-halo scatter expected. After careful consideration of recent simulation results, 
we conclude that the concentrations of the smallest halos, while well-determined at high redshift,  are still uncertain by a factor of 4--6 when extrapolated to low redshift. 
This in turn produces up to two orders of magnitude uncertainty in the predicted annihilation flux for any halo mass above this scale. Substructure evolution, the small-scale cutoff to the power spectrum, cosmology, and baryonic effects all introduce smaller, though cumulative, uncertainties. We then consider intrinsic variations from halo to halo. 
These arise from variations in concentration and substructure, leading to a scatter of $\sim$2.5 in the predicted annihilation luminosity. 
Finally, we consider the problem of detecting gamma rays from annihilation, given the expected contributions from other sources. We estimate the signal-to-noise ratio 
for gamma-ray detection as a function of halo mass, assuming that the isotropic gamma-ray background and cosmic rays from star formation are the main noise sources in the detection. This calculation suggests 
that group-scale halos, individually or in stacks, may be a particularly interesting target for the next generation of annihilation searches.}
\begin{document}
\maketitle
\flushbottom

%section 1
\section{Introduction}
\label{sec:intro}

Dark Matter is a key part of the current cosmological model, producing the rich array of large-scale structure we see around us in the present-day Universe.
Although there is evidence for its existence and detailed distribution from many different astrophysical probes, including the Cosmic Microwave Background
\cite[CMB --][]{Planck2015}, large-scale structure \cite[e.g.,][]{Alam2017}, cosmic shear \cite[e.g.,][]{Joudaki2017}, mass reconstructions of galaxy clusters 
\cite[e.g.,][]{Umetsu2015,Biviano2017,Lagattuta2017}, galaxy-galaxy lensing \cite[e.g.,][]{Leauthaud2012,Velander2014,Hudson2015,vanUitert2016}, 
satellite dynamics around bright galaxies \cite[e.g.,][]{More2011}, field galaxy rotation curves or velocity dispersions \cite[e.g.,][]{Ouellette2017}, and the velocity dispersions of the dwarf satellites of the Local Group \cite[][and references therein]{McConnachie2012}, the fundamental nature of dark matter remains unknown. 
 
Although there are many candidates for the dark matter particle \cite[see e.g.,][for a review]{Feng2010}, Weakly Interacting Massive Particles (WIMPs) remain a favourite candidate, not least because of the `WIMP miracle', that thermal relics with a weak-scale cross-section and a mass in the 100 GeV--1TeV range naturally produce the correct mass density in the present-day Universe \cite{Feng2010, Gaskins2016}. In principle, WIMPs can be detected directly in the lab when they interact with a standard model particle. In practice, these `direct' detection experiments have not yet produced a conclusive discovery, although their limits are improving steadily \cite[see][for reviews]{Snowmass2013,Marrodan2016}. In the mean time, `indirect' searches for WIMP annihilation or decay products provide an important alternative avenue for identifying dark matter. In particular, indirect searches already rule out candidates with a thermal cross-section at the lowest masses, at least for some annihilation channels \cite[e.g.,][]{Ahnen2016}. If this trend progresses up to mass limits around a TeV, the WIMP may become a much less attractive candidate.

Many WIMP candidates are predicted to annihilate into detectable standard model end-states, including neutrinos, electrons/positrons, and gamma rays \cite{Feng2010}. The gamma-ray signature from annihilation is particularly important, as it may be one of the few direct probes of dark matter structure. As a two-body process, the annihilation signal from a given region goes as the local density squared, and thus it depends on the overall inhomogeneity of the particle distribution. Predicting the expected flux requires a detailed model of the dark matter distribution on all scales, and thus represents a challenge for our understanding of structure formation.

A number of systems have been proposed as targets for gamma-ray searches for the dark matter annihilation signal 
\cite[see][for recent reviews]{Charles2016,Gaskins2016}, 
including the Galactic Centre \cite[e.g.,][]{Archer2014,Abdallah2016,Ajello2016}, the centres of other nearby, luminous galaxies 
 \cite{Gondolo1994,Baltz2000,Fornengo2004,Tasitsiomi2004,Mack2008,Saxena2011,Buckley2015,Caputo2016,LiHuang2016}, 
the dwarf satellites of the Milky Way and M31 \cite[e.g.,][]{GeringerSameth2011,Ackermann2011,Ackermann2014,Ackermann2015,Ahnen2016,Albert2017}, 
galaxy clusters \cite[e.g.,][]{Ackermann2010,Arlen2012,Huang2012,Prokhorov2014,Ackermann2015b,Ackermann2016b,Liang2016,Anderson2016,Quincy2016}, 
the smooth extra-galactic background \cite[e.g.,][]{Bergstrom2001,Ullio2002,Taylor2003,DiMauro2015,Ackermann2015c}, 
and large-scale fluctuations present within this background \cite[e.g.,][and many others -- see previously cited reviews for full references]{Ando2009,Camera2013,Ando2014,Shirasaki2014,Camera2015,Cuoco2015}. 
While the signal is predicted to be strongest from the Galactic Centre, other known or possible components in that region obscure it \cite{Abazajian2014,vanEldik2015,Ajello2016}. 
Relative to known components, there is a detected excess in gamma-ray emission from the Galactic Centre \cite{Goodenough2009,Hooper2011,Daylan2016}, 
but it is unclear whether it comes from dark matter annihilation or from a previously unknown population of conventional sources such as pulsars, which can produce a very similar gamma-ray spectrum \cite{Baltz2007}.

Beyond our galaxy, the most constraining targets are generally local dwarf galaxies \cite{Charles2016,Ahnen2016}, considered jointly or in stacks. 
On the other hand, the only direct probe of the mass distribution in these objects, their stellar velocity distribution, is restricted to their innermost regions
\cite[][and references therein]{McConnachie2012}. The correspondence between the known dwarfs and the dense halo substructure predicted by simulations is increasingly well understood \cite{Sawala2016}, but remains complex and slightly model dependent \cite[e.g.,][]{Oman2015}.

Relative to these targets, nearby extra-galactic halos are somewhat less constraining \cite{Sanchez-Conde2011,Charles2016}. On the other hand, they have several other advantages, including angular sizes well-suited to the angular resolution of the Fermi Large Area Telescope (LAT) and ground-based Air Cerenkov Telescopes (ACTs), multiple, independent probes of their mass distributions, and in some cases, locations on the sky well away from contaminating Galactic emission. The correspondence of isolated clusters, groups and galaxies to simulated dark matter halos is relatively uncontroversial, and mean trends in halo mass versus stellar mass have been established empirically, e.g.~using weak gravitational lensing \cite{Leauthaud2012,Velander2014,Hudson2015,vanUitert2016}. Furthermore, as discussed below, predictions of the annihilation flux from 
extra-galactic halos assume typical concentrations and substructure content, but the natural scatter in these properties seen in simulations 
suggests some individual objects may be boosted significantly with respect to the mean. Possible targets of this kind considered recently include 
nearby galaxy clusters \cite[e.g.][]{Arlen2012,Huang2012,Prokhorov2014,Ackermann2015b,Ackermann2016b,Liang2016,Lisanti2017}, 
but also individual galaxies \cite[e.g.][]{Saxena2011,Buckley2015,Caputo2016,LiHuang2016}. 
In particular, an unexpected gamma-ray component has been detected in the centre of M31 \cite{Ackermann2017}; 
whether this can be explained by pulsars or some other conventional population remains to be seen.

In this paper, we reconsider the annihilation signal from extra-galactic halos, in light of recent simulation results on halo density profiles, halo concentration, and halo substructure. We summarize the uncertainties, particularly those in the concentration-mass relation, which remain considerable for reasons discussed below. We also consider the effect of halo-to-halo scatter on the predictions. Finally, given current detections of at least some gamma-ray emission from nearby halos, we estimate the signal-to-noise ratio (SNR) expected from gamma-ray observations, given realistic astrophysical backgrounds. Where needed, we assume 
the Planck 2015 \cite{Planck2015} cosmological parameters: $\Omega_{\rm m} = 0.3089,$ $\Omega_{\rm b} = 0.0486,$ $h = 0.677,$ $n_{\rm s} = 0.967,$ $\Omega_{\Lambda} = 0.6911,$ $\sigma_8 = 0.8159$. 

%section 2
\section{Predicted emission from smooth halos}
\label{sec:boost}

%subsection 2.1
\subsection{The halo boost factor $B_h$}
\label{sec:2.1}

Given pairs of dark matter particles $\chi$ (assumed, for Majorana WIMPs, to be their own antiparticles $\bar{\chi}$) of mass $m_{\chi}$ and density $\rho$ 
in a volume $V$, the rate at which they will annihilate into other particle/antiparticle pairs is 
\begin{equation}
 R = \frac{\langle \sigma v\rangle}{2m_{\chi}^2}\int_V{\rho^2 dV},
\label{eq:ann}
\end{equation}
where  $\langle \sigma v\rangle$ is the velocity-averaged annihilation cross-section.
We can separate the rate into two factors, one depending only the particles' fundamental properties $m_{\chi}$ and $\langle \sigma v\rangle$, and the other depending only on their spatial distribution
\begin{equation}
 R = \left[\frac{\langle \sigma v\rangle}{2m_{2\chi}^2}\right]\left[\bar{\rho}^2 V B(V)\right],
\label{eq:annfactors}
\end{equation}
where we have defined a dimensionless factor
\begin{equation}
B(V) \equiv \frac{1}{\bar{\rho}^2 V}\int{\rho^2 dV}
\label{eq:boost_halo}
\end{equation}
that characterizes the inhomogeneity of the particle distribution. This factor has been variously called the ``enhancement'' \cite{Ullio2002}, the ``flux multiplier'' 
\cite{Taylor2003}, the ``clumpiness'' \cite{Lavalle2008}, or the ``(cosmological) boost factor'' \cite{Cirelli2011}. In what follows we will refer to it as the (halo) boost factor $B_h$, distinguishing this from the (substructure) boost factor $B_{\rm sub}$ discussed in section \ref{sec:substructure}, or the total boost factor including both terms. The boost factor is a function of the volume considered. For halos, we will take this to be the spherical volume bounded by the virial radius, defined to be $r_{\rm vir} = r_{200c}$, the radius within which the mean density is 200 times the critical density of the universe at that redshift, $\rho_c(z)$. (Similarly, halo masses will be taken to be $M_{200c}$, the mass enclosed within $r_{200c}$.) For this choice of volume, we will denote the halo boost factor simply $B_h$.

We can calculate the halo boost factor directly from the spherically-averaged density profile. For the halos seen in cosmological simulations, 
this has most often been approximated as the Navarro, Frenk and White (NFW) profile \citep{NFW96}:
\begin{equation}
\rho_{NFW} = \frac{\rho_s}{(r/r_s)\left[1 + (r/r_s)\right]^2},
\end{equation}
which has free parameters $r_s$ (the scale radius where the logarithmic slope of the profile is $d\ln \rho/d\ln r = -2$), and $\rho_s$, a characteristic density. For a given mass and a definition of the virial radius $r_{\rm vir}$, the profile depends only on the scale radius, or equivalently on the concentration parameter $c\,\equiv\, r_{\rm vir}/r_s$.  
An analytic expression for $B_h(c)$ for the NFW profile is given in, e.g.,~\cite{Taylor2003}.

More recently, high-resolution simulations have found that cosmological halos deviate slightly but systematically from the NFW fit \cite{Navarro2004,Merritt2006,Gao2008}, and that a better approximation, at least over the range of radii currently resolved, is the Einasto profile: 
\begin{equation}
\rho_{\rm Einasto} = \rho_{-2} \exp{\left[-\frac{2}{\alpha}\left\lbrace\left(\frac{r}{r_{-2}}\right)^{\alpha} - 1\right\rbrace\right]}.
\end{equation}
This fit has a characteristic radius $\rho_{-2}$ equivalent to the NFW scale radius, as well as an additional shape parameter $\alpha$ that controls the overall 
curvature of the logarithmic slope. For values of $\alpha\sim0.2$--0.25, the Einasto profile is very similar to the NFW profile, over a broad range in radius. 
For larger values of $\alpha$, the profile becomes flatter in the inner regions of the halo, and steeper in the outer regions. 

For the Einasto fit, the boost for a halo of a given mass and virial radius depends on the two parameters $\rho_{-2}$ and $\alpha$, or equivalently 
on a concentration parameter $c\,\equiv\,r_{\rm vir}/r_{-2}$ and $\alpha$. (There is a closed form analytic expression for $B_h(c,\alpha)$, though it is slightly more complicated than the NFW version.) The two parameters $c$ and $\alpha$ are correlated, however, so the average shape of the profile can be expressed as a single-parameter function of the peak height parameter $\nu \equiv \delta_c/\sigma(M,z)$, where $\delta_c$ is the critical overdensity threshold for collapse and $\sigma(M,z)$ is the r.m.s.~of density fluctuations on mass scale $M$. In terms of $\nu$, the mean value of the shape parameter has been fitted as 
\begin{eqnarray}
\alpha (\nu) &=& 0.155 + (0.0095 \nu^2),\ \rm{or}\ \nonumber \\
\alpha (\nu) &=& 0.115 + (0.014 \nu^2)
\end{eqnarray}
by \cite[][G08 hereafter]{Gao2008} 
and \cite[][K16 hereafter]{Klypin2016} 
respectively. The fit in 
%\cite{Klypin2016} 
K16 was based on more data at high $\nu$, so it seems likely to be more reliable there. The origin of the disagreement at low $\nu$ is unclear; there may be some 
residual redshift dependence, as the low-redshift results generally prefer a higher value of $\alpha$ at low $\nu$. In the absence of further evidence, we will 
assume the more recent results of K16 to be more accurate overall. We also note that values $\alpha > 0.2$ are only seen in high peaks with $\nu \gtrsim 2.5$, which 
generally have low concentrations.

%figure 1
\begin{figure}[t]
\centering
\includegraphics[width=0.9\textwidth]{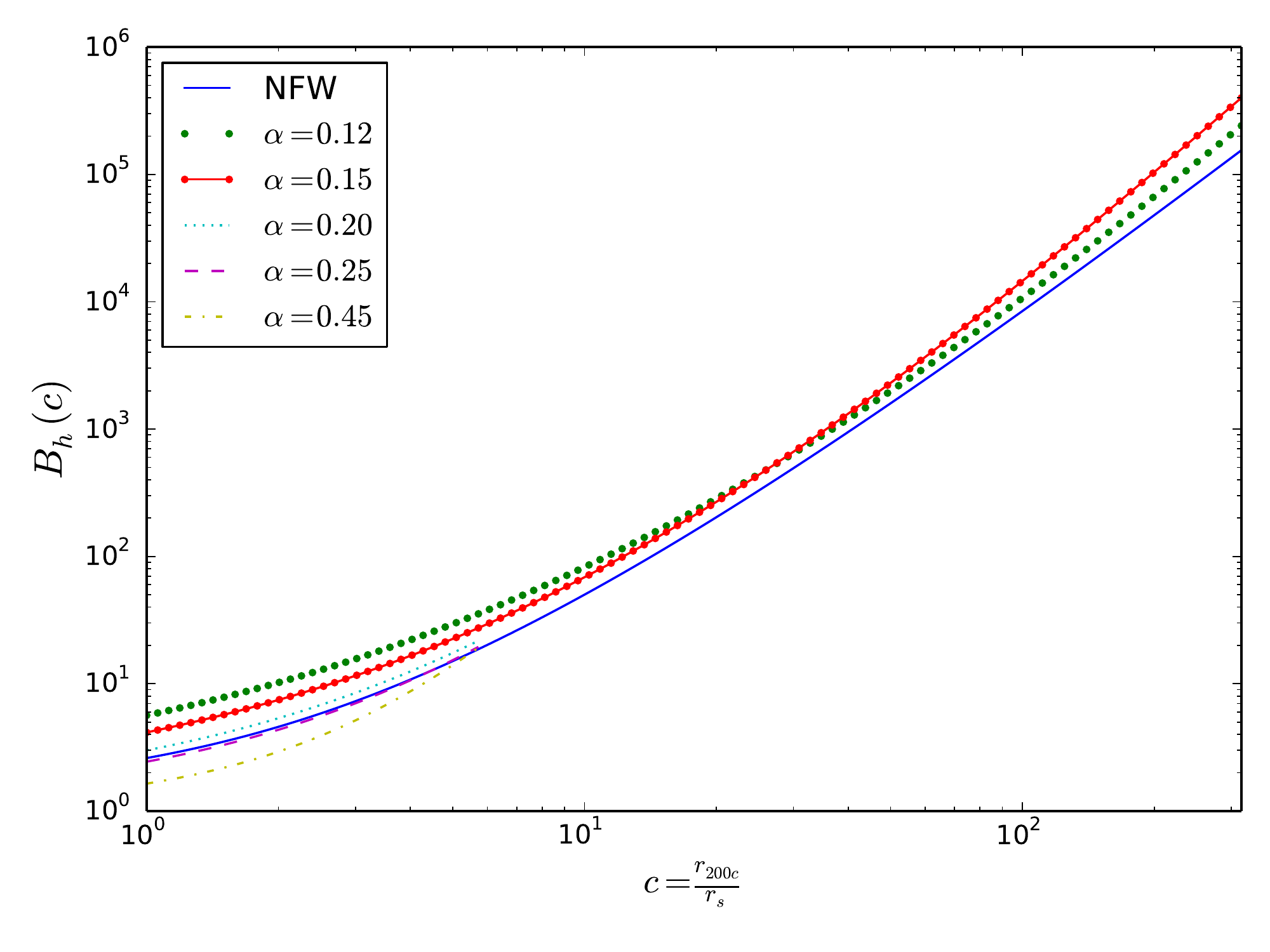}
\caption{The halo boost factor as a function of concentration, for the NFW (solid curve) and Einasto profiles (points and dashed/dotted lines, for different values of the shape parameter as indicated). In the latter case, the boost factor has been plotted only over the range of concentrations expected for field halos with that value of the shape parameter.} 
\label{fig:boost_conc}
\end{figure}
Figure \ref{fig:boost_conc} shows the dependence of $B_h(c)$ on the concentration $c\,\equiv\, r_{\rm vir}/r_s$ for halos with a NFW profile (solid curve), and on the concentration $c\,\equiv\,r_{\rm vir}/r_{-2}$ for Einasto profiles with several different values of the shape parameter $\alpha$ (points/lines). In each case, we have only plotted the curve over the range of concentrations expected for field halos of that shape parameter. Thus, for $\alpha \ge 2$, we only plot the boost for concentrations $c < 6$, since these are the expected values for high-$\nu$ peaks, as discussed in section~\ref{subsubsec:cnu} below.

In general, the halo boost varies slowly for very small concentrations ($c \lesssim 5$), but then increases as a power-law of slope ~2.5--2.8 (i.e.~somewhat shallower than the slope of 3 used in some approximations) at higher $c$. A convenient fit to the exact form is:
\begin{equation}
B_h(c) = A(c+B)^C\,.
\label{eq:Bhfit}
\end{equation}
Over the range $c = 1$--250, this fit with $(A, B, C)  = (0.08,3.0,2.5)$ matches the exact NFW form to better than 5\%, while fits with $(A, B, C)  = $(0.1,4.5,2.5) and (0.04,4.3,2.78) fit the exact Einasto forms for $\alpha$ = 0.115 and 0.155, the low-$\nu$ values in the fitting functions given above, to better than 12\%.

%subsection 2.2
\subsection{Halo concentration}
\label{sec:2.2}

\subsubsection{General relations for large masses}
In the most general sense, concentration characterizes the central density of a system relative to its mean density, or the mass within some small radius relative to the total mass. This is most often described by a concentration parameter $c$ defined in terms of a ratio of radii $r_{\rm vir}/r_s$ or $r_{\rm vir}/r_{-2}$ \cite{NFW96}, as above, but can also be characterized by a ratio of circular velocities, e.g. the ratio of the peak circular velocity to the velocity at the virial radius $v_{peak}/v_{\rm vir}$ \cite{Prada2012,Klypin2016}, or by a ratio of densities, e.g.~$\rho_{peak}/\rho_{c}$, \cite{Alam2002,Diemand2007,Moline2017}. For a given fit to the profile, there is a fixed relationship between these different concentration parameters, but the different definitions can introduce systematic differences in concentration when the halo profile is fit with different functional forms, or if the shape of the profile varies systematically with mass or redshift. Here we will use the first definition by default, with the outer boundary $r_{\rm vir} = r_{200c}$, and relate literature results to this definition. Also we note that the halo-to-halo scatter in concentration is considerable, as discussed further below. Thus we will generally be interested in the mean or median value for a sample of halos covering some range of mass and redshift.

The mean relation between halo concentration, mass and redshift has been studied extensively in simulations \cite[e.g.][]{NFW96, Bullock2001,Maccio2008,Zhao09,Klypin2011,Prada2012,Dutton2014,Diemer2015,Correa2015,Klypin2016} -- see \cite{Moline2017} for further references. Several general trends are well-established. Recently-formed halos have concentrations of $c\sim3$--4; as their growth rate slows and major mergers become less frequent, their central density profile and scale radius remain roughly constant, while their virial radius increases as $\rho_c$ drops, leading their concentration to increase as roughly $(1+z)^{-1}$ \cite[e.g.,][]{Zhao09}. Thus, concentration is an indicator of the change in critical density since the core of the halo formed; higher-mass halos considered at a given redshift, or halos of a given mass considered at a higher redshift, have generally formed more recently in relative terms and therefore have lower concentrations, down to a minimum value of $c\sim3$.

Beyond these general trends, several controversies exist. Contrary to expectations, some recent concentration-mass relations have predicted an {\it increase} in concentration with mass and/or redshift, when the mass is much larger than the typical collapse mass for that redshift 
\cite{Klypin2011, Prada2012,Dutton2014, Diemer2015}, while others have disagreed, claiming these trends are due to non-equilibrium effects \cite[e.g.][]{Ludlow2012}. Most recently, work by Klypin et al.~\cite{Klypin2016} has clarified that relaxation issues aside, the most massive halos are genuinely more concentrated, possibly due to more radial infall patterns from the surrounding density field, but also differ significantly from NFW profiles. Part of the disagreement over the concentration-mass-redshift relation comes from comparing concentrations based on radial ratios to concentrations based on velocity ratios, for halos fitted alternately with NFW profiles or Einasto profiles. While a concentration parameter based on velocity ratios does increase at the highest masses, the ratio of the virial radius to the Einasto scale radius $r_{-2}$ remains roughly constant and equal to $\sim3$. Thus, this controversy seems mostly resolved (and will also be less relevant here, since it affects only the most massive halos, none of which are located in the very local volume (at distances D $\lesssim10$ Mpc) considered below.) 

Another significant disagreement relates to extrapolation of the concentration-mass relation to very low halo masses. Simulations of cosmological volumes can typically
resolve halo growth over only a few decades in mass below the characteristic cosmological mass scale $M_*$ for which $\nu =1$. Over this relatively short 
baseline, the mean concentration-mass relation is reasonably well fit by a power-law. Extrapolating such power-law fits to the smallest halo masses expected 
for canonical WIMP models predicts extremely large concentrations, and thus very large annihilation rates \cite[e.g.,][]{Pinzke2011,Gao2012}. There is no physical 
basis for this extrapolation, however;  the trend in concentration with mass is due to the variation in formation epoch with mass, which should flatten at low masses. 
Simulations at smaller mass scales and higher redshifts have demonstrated that models based on the r.m.s.~fluctuation amplitude $\sigma(M)$ or the peak height 
$\nu$ are both more physical and a better fit to the data \cite{Prada2012,Sanchez-Conde2014}, while simulations in different cosmologies have shown that they 
are also more universal, and less dependent on the specific value of cosmological parameters \cite[e.g.,][]{Diemer2015}.  

So far we have discussed fits to the concentration-mass-redshift relation as determined empirically from simulations. Recently, \cite[][OA16 hereafter]{Okoli2016} 
proposed a theoretical concentration-mass relation, derived by considering the ellipsoidal collapse of perturbations, and the conservation of energy of the collapsing 
region. Their model compares quite well with the features seen in simulations at $z=0$, and also in higher-redshift simulations on smaller mass scales. 
Thus, we will also consider this model, and in particular, its predictions for very low-mass halos.

\begin{figure}[t]
\centering
\includegraphics[width=0.9\textwidth]{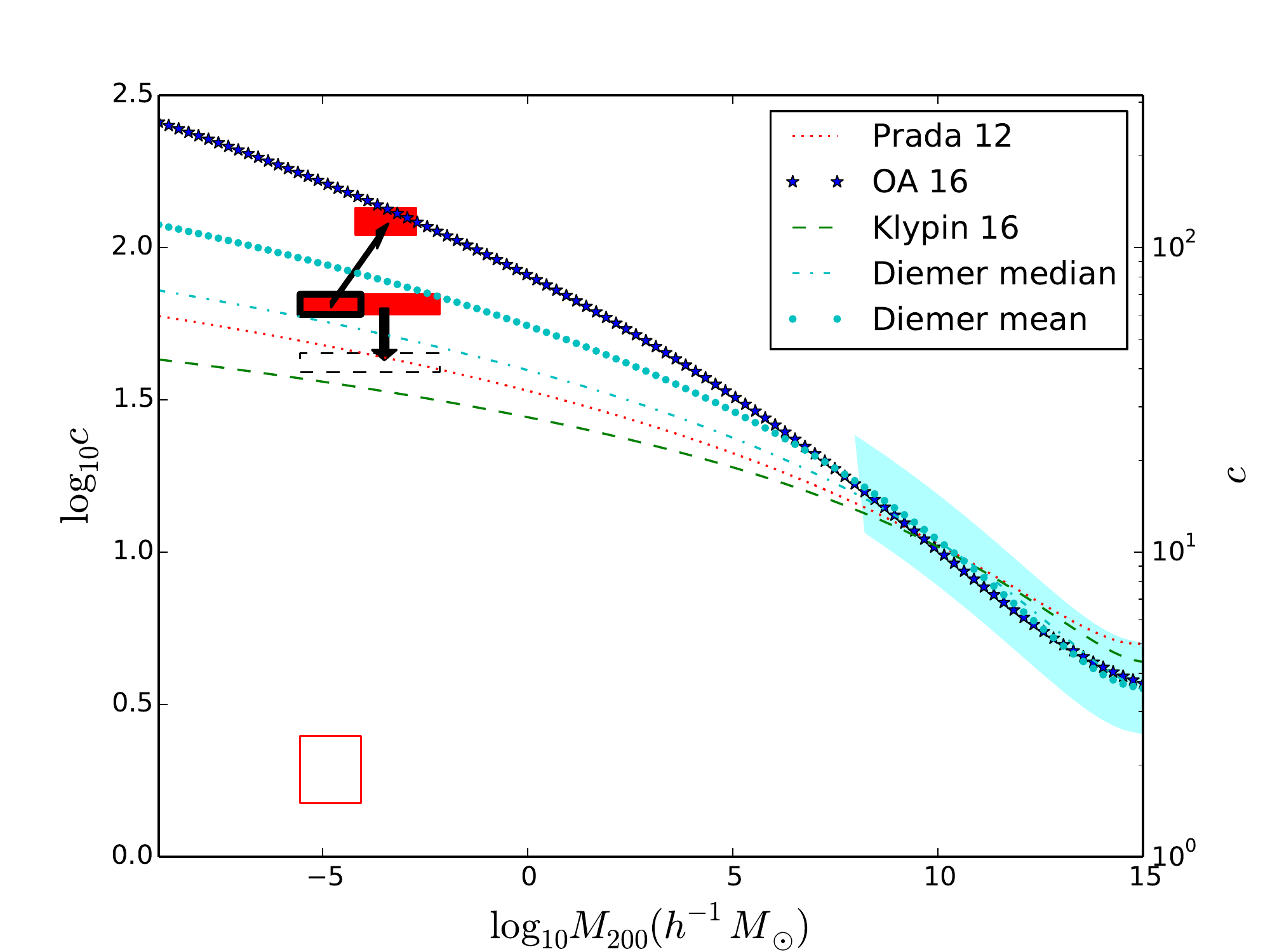}
\caption{Concentration-mass relations at  $z = 0$, from \cite[][-- labelled `Klypin 16',`Prada 12', `Diemer median' and `OA16' respectively]{Prada2012,Diemer2015,Klypin2016,Okoli2016}. The cyan region shows the range of halo mass sampled by the 
simulations on which the numerical fits of \cite{Prada2012,Diemer2015,Klypin2016}  were based. The vertical range of the cyan region indicates the mean dispersion in the concentration of the
simulated halo samples. The long solid rectangle indicates the range of mass and mean concentration from the lowest redshift output of \cite{Ishiyama2014}, extrapolated 
to the present day using the redshift scaling $c \propto (1+z)^{-1}$ originally proposed by \cite{Bullock2001}. The thick arrow and lower, dashed rectangle show how 
these concentrations change using the alternate redshift scaling  $\propto \rho_c^{-1/3}$, recently proposed by \cite{Pilipenko2017}. The lower open square at $\log_{10} c\sim0.3$ indicates the 
range of mass and concentration measured by  \cite{Ishiyama2014} for three individual halo profiles at $z=32$. The thick open rectangle at $\log_{10} c\sim1.8$ 
shows these quantities 
extrapolated to $z=0$ assuming $c \propto (1+z)^{-1}$, while the arrow pointing up and to the right and the upper solid rectangle show the predicted mass and 
concentration for these systems at $z=0$ if a central-density conserving model is assumed instead.} 
\label{fig:conc_range}
\end{figure}

Figure \ref{fig:conc_range} shows several recent predictions of concentration as a function of halo mass at $z = 0$, in our adopted Planck 2015 cosmology. 
The models shown are the median relations from \cite[][-- labelled `Klypin 16',`Prada 12', `Diemer median' and `OA16' respectively]{Prada2012,Diemer2015,Klypin2016,Okoli2016}, as well as the mean relation from \cite{Diemer2015}. 
As discussed further below, the distribution of concentration in a given halo mass bin is approximately log-normal \cite[though not quite -- see][]{Diemer2015}, 
so the offset between the median and median values is expected. Comparing results for the median relation, we find that the three different fits to simulation 
data at low redshift \cite{Prada2012,Diemer2015,Klypin2016} agree to better than 0.1 dex over the range of halo mass directly probed by the simulations they 
are based on, highlighted in cyan on the plot. 

%figure 2

\subsubsection{Extrapolation to very small masses}

To calculate the halo boost factor, we need to understand  the cold dark matter (CDM) density distribution on all scales, down to the limit where it becomes smooth. For conventional WIMPs, this lower mass scale is set by free-streaming and/or acoustic oscillations after kinetic decoupling from Standard Model particles in the early universe, and can range from $\sim10^{-12} M_\odot$ to $10^{-3} M_\odot$, depending on the mass and couplings of the WIMP, while for other dark matter candidates it can be even smaller \cite[e.g.][and references therein]{Bringmann2009}. The value $10^{-6} M_\odot$ (corresponding to a standard 100 GeV WIMP) is sometimes taken as a fiducial reference 
point, as in the models of \cite[][]{Diemand2005,Moline2017}.
Extrapolating the fitting formulae for the concentration-mass relation derived in low-redshift, large-volume numerical studies \cite{Prada2012,Diemer2015,Klypin2016} 
down to these very small mass scales, we find that they diverge slightly,  but are still consistent at the 0.2 dex level down to the smallest masses considered here, as 
shown in figure \ref{fig:conc_range}. The analytic model of OA16,
%\cite{Okoli2016}
based on energy conservation and the ellipsoidal collapse model, agrees with the simulation results over the range where they 
have been tested directly. At these very low masses, however, it predicts concentrations up to $\sim6$ times higher. 

Given the analytic model makes several simplifying assumptions, it seems possible a priori that the empirical models calibrated off simulations should be more reliable; 
in particular, several authors have claimed that high-redshift simulations of very small mass scales directly confirm the extrapolations shown in figure \ref{fig:conc_range} \cite{Sanchez-Conde2014,Moline2017}. It is worth reconsidering carefully, however, the evidence for lower concentrations in {\it present-day} systems of very low mass.

Several numerical studies have directly simulated the formation of `microhalos', the smallest structures to form in a cold dark matter cosmology with conventional WIMP dark matter \cite{Diemand2005,Ishiyama2010,Anderhalden2013,Ishiyama2014}. In each case, the authors measured concentrations of $c\sim2$ for these low-mass halos, {\it at high redshift} when the halos had recently formed. Given the mass scale corresponding to the entire simulated region would go non-linear long before $z=0$, they were not able to trace the fate of these structures directly down to low redshift. Thus, these simulations give direct evidence for the profile shapes, concentrations and densities expected at $z\gtrsim\,$25--30, but not at redshifts below this. 

To extrapolate the high-redshift results down to $z=0$,  previous authors \cite{Sanchez-Conde2014,Moline2017} have used the mean redshift dependence of the concentration-mass model of, e.g. \cite{Bullock2001}, $c \propto (1+z)^{-1}$. With this scaling, the predicted concentrations at $z=0$ would be $c \sim 60$--70, consistent 
with their earlier predictions \cite{Prada2012}. These extrapolated values are illustrated by the long solid rectangle on the figure, where we have taken the masses, 
concentration values, and scatter from the lowest redshift output of \cite{Ishiyama2014}, the most recent and largest-volume set of microhalo simulations. 

More recently, \cite{Pilipenko2017} have pointed out that a more accurate scaling with redshift should in fact be $c \propto \rho_c^{-1/3} \propto [H(z)/H_0]^{-2/3}$, where 
$H(z)$ is the Hubble parameter. The implicit assumption here is that since concentration is a ratio between a virial radius $r_{\rm vir}$ and a scale radius $r_s$, if the latter 
remains fixed, then concentration will scale with virial radius. Assuming the virial radius is defined by an overdensity criterion, then in the absence of a significant change 
in halo mass, it will evolve with redshift as  $\rho_c^{-1/3}$. This redshift scaling predicts even lower concentrations for low-mass halos at $z=0$, roughly 0.1 dex below the lowest relations plotted in figure \ref{fig:conc_range},  as indicated by the long dashed rectangle and the downward-pointing arrow. 

This redshift scaling assumes, however, that the mass inside the virial radius remains constant, even as this radius increases by a factor of $\sim20$--30 or more. Thus, 
it assumes the unrealistic situation where a halo forms at high redshift in a void, with no significant mass outside its (high-redshift) virial radius $r_{\rm vir} \sim 2$--5$\,r_s$. 
In practice, the density distribution around cosmological halos always extends smoothly well beyond the virial radius. Assuming this remains true at high redshift, as the 
virial radius defined by the overdensity criterion increases, it will enclose more matter; the resulting increase in halo mass will produce an additional increase in the virial 
radius relative to the scale radius, and thus concentration will grow {\it faster} than $\rho_c(z)^{-1/3}$ or $(1+z)^{-1}$. 

We can make a simple estimate of the final concentration for the lowest-mass systems by assuming that their density structure within $r_s$ remains {\it constant with 
redshift}, and that the profile is NFW out to the virial radius at $z=0$. In this case, the mass will evolve as $M(z) \propto m(c(z)) = \ln(1+c) - c/(1+c)$, the virial radius 
$r_{\rm vir}$ will increase as $[M(z)/\rho_c(z)]^{1/3}$, and the concentration will increase as $r_{\rm vir}$. This `central density conserving' model thus predicts both 
a larger final mass at redshift $z=0$, and a larger concentration than the two other extrapolations. The resulting shift in mass and concentration, relative to values 
scaled by $(1+z)^{-1}$, is indicated by the arrow pointing up and to the right on figure~\ref{fig:conc_range}, while the upper solid rectangle shows the predicted 
masses and concentrations at $z=0$. We note that this extrapolation agrees almost exactly with the model of OA16,
%\cite{Okoli2016}, 
suggesting that this analytic model may be a realistic description of CDM structure on the smallest scales. 

\begin{figure}[t]
\centering
\includegraphics[width=0.9\textwidth]{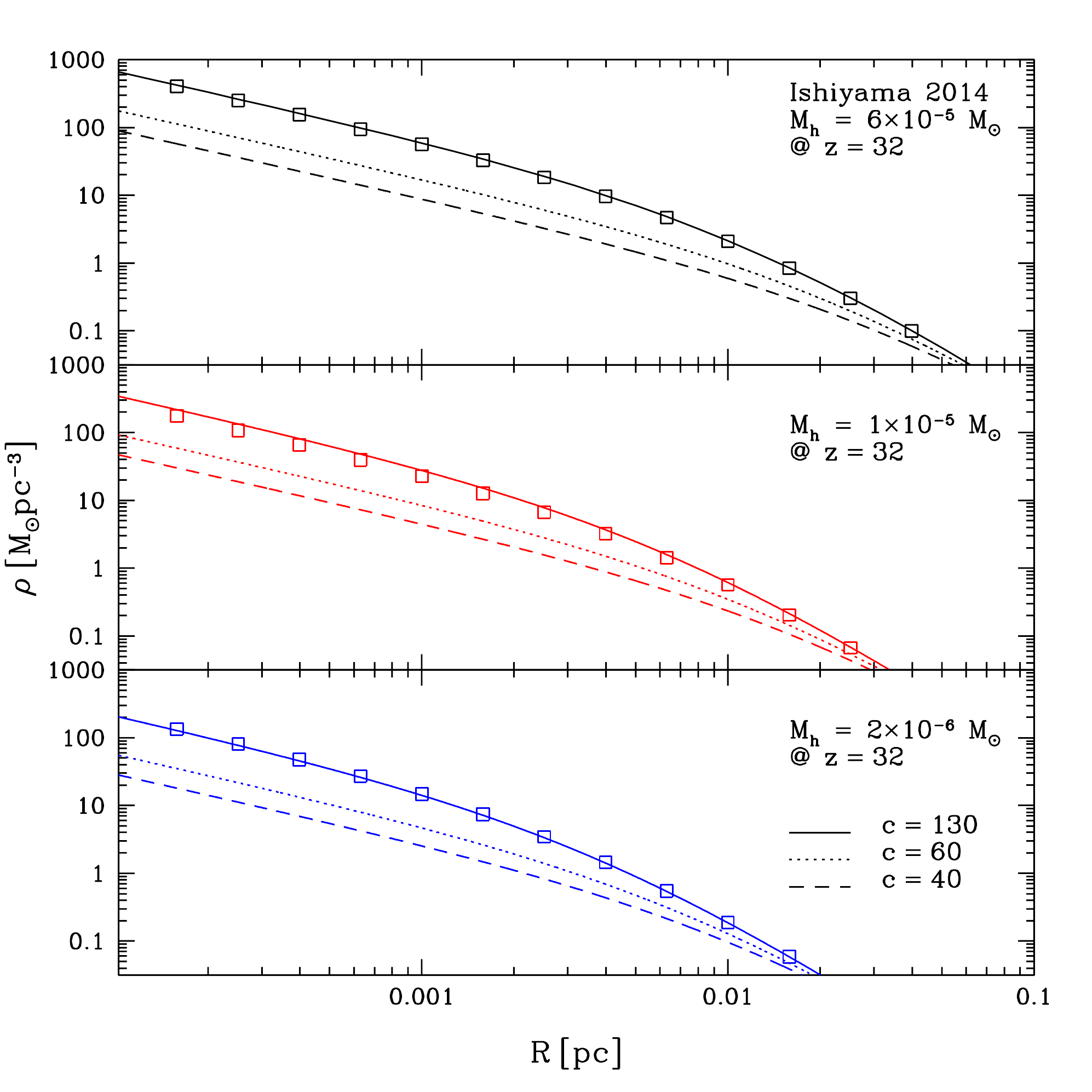}
\caption{Density profiles measured by \cite{Ishiyama2014} at $z=32$ for three individual halos, with the masses $M_h$ listed (open squares). The curves show density profiles for halos of different concentrations at $z=0$. For each of these, the total halo mass has been extrapolated to $z=0$ as explained in the text, and is roughly 9 times the value at $z=32$. In the absence of density evolution, the profiles are consistent with $z=0$ concentrations of $c\sim130$, not $c\sim$40--60.}  
\label{fig:profilecomp}
\end{figure}

We can also illustrate the disagreement between different models for the redshift evolution of concentration by comparing the predicted density profiles directly. 
Figure~\ref{fig:profilecomp} shows density profiles at $z=32$ for three individual halos from the simulations of \cite{Ishiyama2014} (open squares). The masses of 
the halos at $z=32$, $6\times10^{-5}M_\odot$, $1\times10^{-5}M_\odot$,  and $2\times10^{-6}M_\odot$, are indicated on the plot; we will assume that these are 
masses within a virial radius of $r_{\rm vir}\sim 2r_s$, since $c\sim2$ is the typical concentration measured at this redshift in their default model with no cutoff to the 
power spectrum. 
These three halos and their surrounding regions were selected at high redshift, and evolved in isolation down to $z=0$. The evolution in isolation demonstrates 
that the halos assembled by $z=32$ can be 
dynamically stable, but is not necessarily realistic in a cosmological context, so we will not consider it further here. The profiles of the haloes at $z=32$ are interesting 
in and of themselves, however. If we extend these profiles out in radius, assuming an NFW profile, we reach 200 times the present-day critical density at virial radii 
of 1.65 pc, 0.91 pc and 0.54 pc respectively. Comparing these values to their high-redshift scale radii (0.013, 0.008, and 0.004 pc respectively), we predict $z=0$ concentrations of 110--135, 80\% higher than those predicted by \cite{Prada2012} or \cite{Diemer2015}. The mass enclosed within the virial radius has increased 
by a factor of $\sim9$ between 
$z=32$ and $z=0$, explaining the larger virial radii and concentrations than a naive scaling as $(1+z)^{-1}$ or $\rho_c(z)^{-1/3}$ would predict. The curves 
on figure~\ref{fig:profilecomp} show the central density structure for halos with masses equal to these extrapolated values, and present-day concentrations of $c=40$, 60 and 130 (dashed, dotted and solid lines respectively). Clearly,  if the central density structure is conserved going from high redshift to low redshift, the simulation results
are consistent with $z=0$ concentrations of $c\sim130$, not $c\sim$40--60. On figure  \ref{fig:conc_range}, we have plotted these masses and concentrations assuming 
a $(1+z)^{-1}$ scaling (thick open rectangle), and the extrapolated mass and concentration assuming density concentration (upper solid rectangle).

%figure 3

We note that the assumption of central density conservation may not be justified. On larger mass scales, simulations have confirmed that the mean 
concentration does evolve as $(1+z)^{-1}$ on average, indicating that central density must gradually drop in typical objects. One possible mechanism for decreasing 
central density is violent merger activity; \cite{Moore2004} have shown that major mergers between similar halos may reduce the concentration by a factor 
of 2 in some cases. Clearly most halos well above the minimum mass scale will have undergone major mergers at some point in their past, explaining why 
the evolution of concentration with redshift is slower than predicted by a density-conserving model. It remains unclear, however, whether this evolution should 
apply to the lowest-mass halos surviving at the present day, which by definition have not accreted much mass over their lifetime, nor to the smallest subhalos, 
which would have been subsumed into more massive systems early on, before they themselves had grown much in mass.

\subsubsection{Concentration and boost factor versus peak height}
\label{subsubsec:cnu}
Given the uncertainties in halo concentration discussed above, we can compare results 
for two possible concentration relations that bracket the likely range behaviour at low mass, 
the model extrapolated from low-redshift, high-mass simulations by K16,
%\cite{Klypin2016}, 
and the analytic model of OA16.
%\cite{Okoli2016}, 
For an NFW profile, these are given by
\begin{equation}
c_{\rm NFW} = 0.522\left[1 + 7.37\left(\frac{\sigma}{0.95}\right)^{3/4}\right]\left[1 + 0.14 \left(\frac{\sigma}{0.95}\right)^{-2}\right]
\label{c_klypin}
\end{equation}
and 
\begin{equation}
c_{\rm NFW} = \left[3.2 + \left(\frac{0.696}{\nu}\right)^{2.32} + \left(\frac{1.71}{\nu}\right)^{1.31} \right]
\label{c_oa}
\end{equation}
respectively. Similar equations may be written for the concentration of haloes with an Einasto profile
\begin{equation}
c_{\rm EINASTO} = 6.5 \nu^{-1.6} \left(1 + 0.21 \nu^2\right)
\end{equation}
\begin{equation}
c_{\rm EINASTO} = 2.28 + \left(\frac{3.25}{\nu}\right)^{1.38}\, ,
\end{equation}
once again for K16
and OA16
respectively. 

\begin{figure}[t]
\centering
\includegraphics[width=1.0\textwidth]{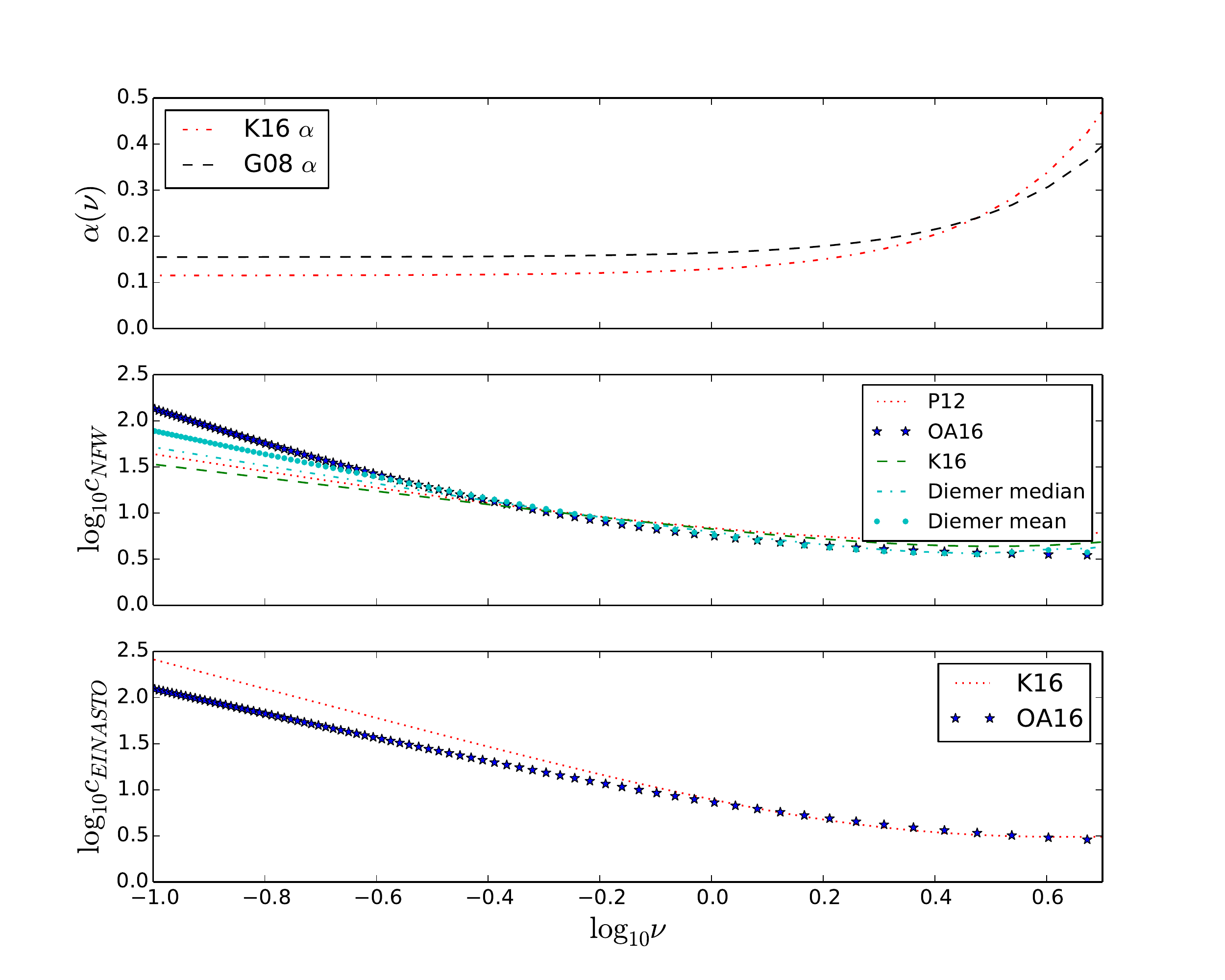}
\caption{(Top panel) Einasto shape parameter $\alpha$ versus peak height $\nu$ from the fits of G08
and K16.
(Middle panel) concentration versus peak height for NFW profiles. (Bottom panel) concentration versus peak height for Einasto profiles. Labels are as in figure \ref{fig:conc_range}.}
\label{c_alpha_nu}
\end{figure}

In the case of the Einasto profile, we also need to specify how the shape parameter $\alpha$ varies with mass and redshift, and/or peak height $\nu$. The top, middle and bottom panels of figure~\ref{c_alpha_nu} show the two fits for $\alpha$ discussed previously in section~\ref{sec:2.1}, the predicted concentration for NFW profiles, and the predicted concentration for Einasto profiles respectively, each as a function of peak height $\nu$. We note that while the OA16 and K16 results provide upper and lower estimates of concentration for NFW profiles, for Einasto profiles their order is inverted. The K16 fit to $c(\nu)$ for Einasto profiles is based on a fairly short baseline in $\nu$, however ($\nu \ge -0.2$), so it may be less reliable for very low peak heights.

%figure 4

\begin{figure}[t]
\centering
\includegraphics[width=0.9\textwidth]{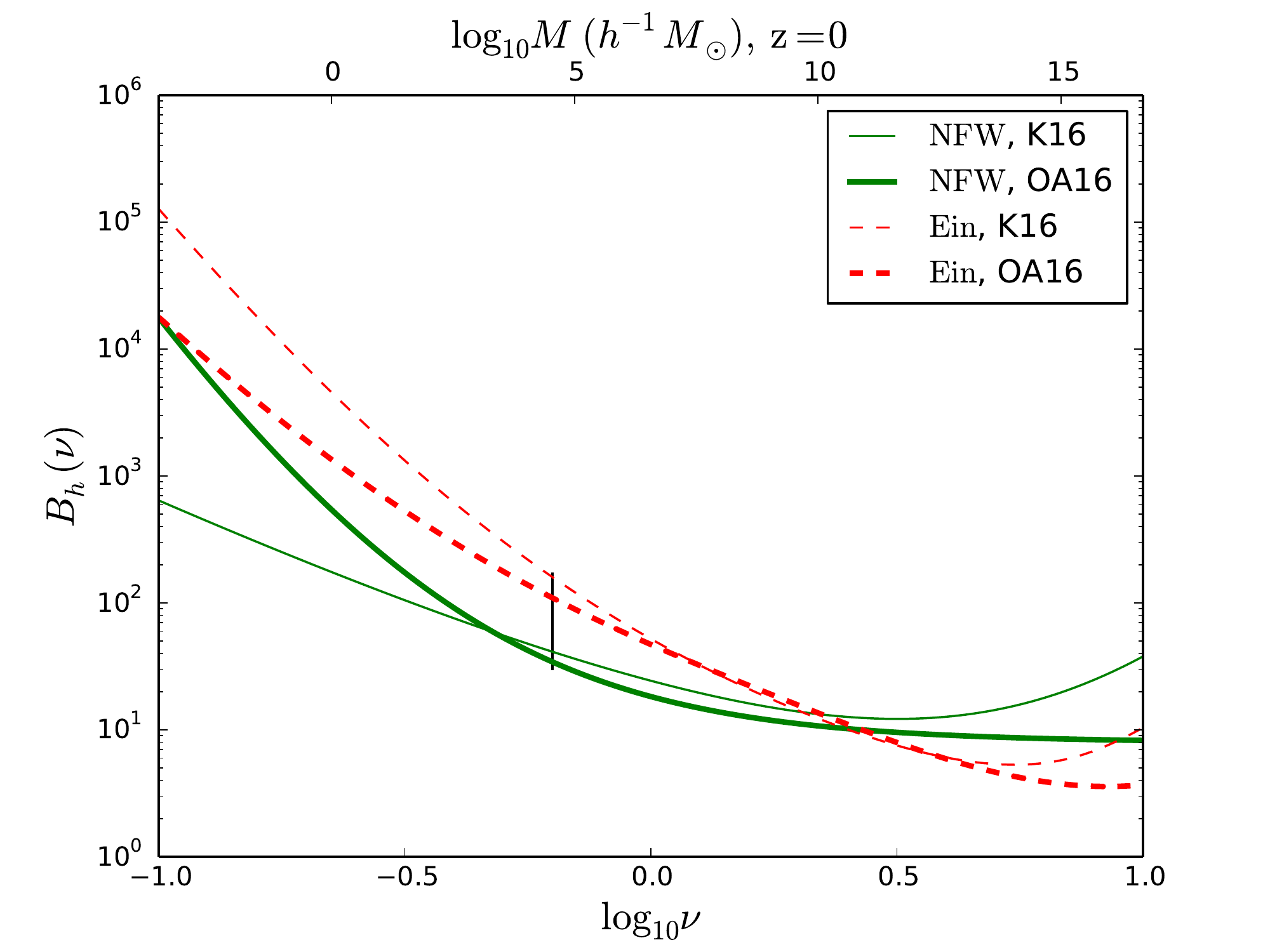}
\caption{The halo boost factor as a function of peak height $\nu$ at $z=0$, for different choices of the density profile and concentration model. 
The solid lines are for the NFW profile, and the dashed lines are for the Einasto profile with the $\alpha$-$\nu$ relation of K16. In each case, 
thick lines indicate the concentration model of OA16, while thin lines indicate the model of K16. The vertical bar indicates the lowest values of $\nu$ simulated by K16.} 
\label{fig:boost_nu}
\end{figure}

Finally, we can combine the halo boost-concentration relation $B_h(c)$ (figure~\ref{fig:boost_conc}) with the concentration-$\nu$ relation $c(\nu)$
to calculate the boost for a smooth halo of typical concentration as a function of peak height, $B_h(\nu)$. Figure~\ref{fig:boost_nu} shows this prediction for two different 
concentration-$\nu$ relations and two different density profiles. Over most of the range in peak height, the scatter between the predictions is about an order of magnitude 
. Only for very low peaks (i.e.~low-mass halos at low redshift) does the range increase to two orders of magnitude.
Throughout the rest of the paper the most conservative possibility, an NFW profile with the K16 concentration model, will be assumed for the boost factor by default, 
although it seems likely that this underestimates the boost factor at low $\nu$ considerably.

%figure 5

\subsection{Scatter in halo concentration}
\label{sec:2.3}

The boost estimates of the previous section assumed the mean halo concentration at a given mass. In fact, the distribution of concentration at fixed mass is approximately, though not quite, log-normal \cite{Bullock2001,Diemer2015}. Since halo boost scales as concentration to a power $C$ between 2.5 and 2.8 (equation~\ref{eq:Bhfit})
we might expect the scatter in concentration to introduce both a larger scatter in the boost, and a bias in the mean value.

To estimate the scatter in the boost factor, we assume a log-normal distribution of concentration, with a 68\%\ r.m.s. scatter in $\log_{10}(c)$ of 0.16 dex, as in the recent measurements of \cite{Diemer2015}. For concentrations $c \gtrsim 10$, the exponent in the boost-concentration relation is $\sim$2.5--2.8, and thus we expect 
0.4--0.45 dex scatter in the boost factor (or a factor of $\sim2.5$--3). We can also estimate the bias in the mean (that is the mean of the boost relative to the boost of the mean) by propagating the scatter through the analytic fit to $B_h(c)$ given previously (equation~\ref{eq:Bhfit}). For the parameters (A, B, C)  = (0.08, 3.0, 2.5), we find that the mean boost is $\sim 20$\%, 36\% and 47\% higher than the boost of the mean for mean concentrations $c=3$, 10, and 30 respectively. Thus, while this bias is small compared to the other uncertainties discussed previously, it is not completely negligible.

% section 3
\section{Effects of substructure}
\label{sec:substructure}

So far we have considered the boost factor for smooth, spherical halos. In fact, cosmological simulations show that the hierarchical merging process through which halos assemble is relatively inefficient, and that the dense cores of smaller systems survive as bound substructure within a halo. Thus, to predict the overall annihilation rate accurately, we need to calculate the extent to which clustering {\it within} a halo increases the integral of $\rho^2dV$. This correction to the smooth halo value is itself 
often called the `boost factor' \cite[e.g.][]{Han2016}; here we will refer to it as the `substructure boost factor', to distinguish it from the (smooth) halo boost factor defined above.

There are several different approaches to calculating the substructure boost factor. A common `halo-based' approach \cite[e.g.][]{Bartels2015,Han2016} is to model the boost from individual subhalos, as well as the distribution of subhalos as a function of radius, infall mass, and degree of tidal stripping. We will present our own version of this calculation below, and show that it agrees well with other recent estimates, \cite[e.g.][]{Bartels2015,Han2016,Stref2017,Moline2017,Hutten2018}. An alternative approach is to model the clustering or density distribution of dark matter particles at a statistical level, calibrating to simulations \cite[e.g.][]{Kamionkowski2008,Kamionkowski2010}. \cite{P2SAD} presents a version of this calculation, based on clustering in phase-space.  
We will consider this approach first, and show it also produces fairly consistent estimates of the additional boost from substructure.

%subsection 3.1
\subsection{$P^2SAD$ prescription}
First, we will investigate the effects of substructure using the particle phase-space average density ($P^2SAD$) model of \cite{P2SAD}. This model is based on the stable clustering hypothesis in phase-space, which assumes that for very small separations in phase-space coordinates $\Delta x$ and $\Delta v$, the average 
number of particles within a phase-space volume does not change with time.  \cite{P2SAD} showed that this hypothesis, together with a tidal stripping model, 
 successfully described the survival fraction and spatial distribution of subhaloes.  Thus, the contribution of substructure (including subhaloes within subhaloes, and 
 all further levels of the hierarchy) to the boost factor may be estimated for various masses and redshifts. The annihilation rate in substructure is given by 
\begin{equation}
R_{\rm sub} = \frac{8\pi^{1/2} b^3}{9 \delta_c^3} 200 \rho_{c,0}f_{\rm sub} M_{200}\frac{\langle \sigma v\rangle}{2m_{\chi}^2}\int_{m_{\rm min}}^{m_{\rm max}}{\mu(m_{\rm col}) m_{\rm col}^{-2}d[m_{\rm col}^2\sigma^3(m_{\rm col})]},
\label{eq:sub}
\end{equation} 
where $f_{\rm sub}$ is the mass fraction in substructure, $\mu(m_{\rm col})$ is the mean fraction of particles that remain bound to a subhalo of mass $m_{\rm col}$ 
that collapsed earlier into a larger structure, 
$m_{\rm col}$, $\delta_c$ is the spherical collapse density, $b = 3.53$, $m_{\rm min}$ and $m_{\rm max}$ are the minimum and maximum masses in substructure 
for a given halo mass, taken to be $10^{-6} M_\odot$ and $\sim 5\%$ of the main halo mass respectively, 
and $\sigma(m)$ is the cosmological variance in density perturbations for a given mass. The substructure fraction $f_{\rm sub}$ is estimated from the subhalo mass function of \cite{2011MNRAS.410.2309G}. For a given host halo, the substructure contribution is calculated by integrating all masses lower than $\sim 5\%$ of the host halo mass, down to $m_{min}$.
Relative to the rate for a uniform-density halo of mass $M_{200}$, $R_{\rm unif} = R_h/B_h$ we can define a substructure boost factor 
\begin{equation}
B_{\rm sub} \equiv \frac{R_{\rm sub}}{R_{\rm unif}} = \frac{8\pi^{1/2} b^3}{9 \delta_c^3} f_{\rm sub}\int_{m_{\rm min}}^{m_{\rm max}}{\mu(m_{\rm col}) m_{\rm col}^{-2}d[m_{\rm col}^2\sigma^3(m_{\rm col})]}\,.
\end{equation}
The total boost factor is then
\begin{equation}
B_{t} = B_h + B_{\rm sub} = B_h \left(1 + \frac{B_{\rm sub}}{B_{h}}\right)\,.
\end{equation}
The dot-dashed curve on figure \ref{boost_substruct} shows ($B_{\rm sub}/B_h$), the relative enhancement from substructure (or `substructure boost'), as a function of halo mass. 

%subsection 3.2
\subsection{Analytic substructure model} 
In this section, we derive an analytic estimate for the extra boost factor from substructure, based on modelling the growth of a halo through mergers, and the evolution of individual subhalos after they merge. Our approach is similar to several other analytic models developed previously \cite[see, e.g.,][for recent examples]{Han2016,Moline2017}, but includes the information on the mass accretion history of an individual halo, and thus allows us to calculate the expected halo-to-halo scatter.

First, we can derive an expression for the relative contribution to the substructure boost factor made by a single subhalo. Consider a main halo of mass $M$ and mean density $\bar{\rho}$ occupying a volume $V$, consisting of a smooth component of mass $M_0$ and mean density $\rho_0$ occupying a volume $V_0$, and a single subhalo of mass $M_1$ and mean density $\rho_1$  occupying a volume $V_1$. The total boost over $V$, $B(V)$ can be broken into contributions from $V_0$ and $V_1$
\begin{eqnarray}
B(V) &=& \left[\left(\bar{\rho}_0\over \bar{\rho}\right)^2 \left(V_0\over V\right) B(V_0) + \left(\bar{\rho}_1\over \bar{\rho}\right)^2 \left(V_1\over V\right) B(V_1)\right]\\
&=& \left[\left(\bar{\rho}_0\over \bar{\rho}\right) \left(M_0\over M\right) B(V_0) + \left(\bar{\rho}_1\over \bar{\rho}\right) \left(M_1\over M\right) B(V_1)\right]\, \nonumber,
\end{eqnarray}
where $B(V_0)$ and $B(V_1)$ are the boosts from the smooth component and the subhalo component respectively. Provided $M_1 \ll M_0$ and $V_1 \ll V_0$, the mean density of the smooth component will be approximately the same as the mean density of the whole halo, $\bar{\rho}_0  \approx \bar{\rho}$. Defining the substructure mass fraction $X = M_1/M$, then the fraction in the smooth component is $(1-X)$. Thus the total boost may be written as 
\begin{equation}
B(V) =  (1-X) B(V_0) + \left(\bar{\rho}_1\over \bar{\rho}\right) X B(V_1) \simeq (1-X) B_h + \left(\bar{\rho}_1\over \bar{\rho}\right) X B(V_1)\,.
\label{eqn:sub}
\end{equation}
We can easily generalize this result to the case of $N$ subhalos
\begin{equation}
B(V) =  (1-X) B_h + \sum_{i=1}^N \left(\bar{\rho}_i\over \bar{\rho}\right) X_i B(V_i) \,.
\label{eqn:sub1}
\end{equation}
where $\bar{\rho}_i$, $X_i$, and $B(V_i)$ are the mean density, mass faction and boost factor of subhalo $i$, occupying volume $V_i$.

To calculate these quantities, we need to define the volumes $V$ and $V_i$ of the main halo and subhalos. 
While the properties of the main halo are calculated within its (cosmological) virial radius $r_{\rm vir} = r_{200c}$, subhalos will be stripped down to 
a smaller tidal radius $r_{t}$ as they orbit. Physically, this radius denotes the average distance at which the tidal force from the host halo becomes 
equal to or greater than the self-gravity of the subhalo. Thus, we need expressions for the subhalo properties -- $X_i = M_i/M, \bar{\rho}_i$, and $B(V_i)$ -- 
within this radius. We explain each of the calculations below.

\begin{enumerate}
\item {Bound mass, tidal radius, and subhalo boost factor}

A subhalo with an initial mass $M_{0}$ and radius $r_{\rm vir}$ at redshift $z_{\rm inf}$ when it falls into the host will be tidally stripped down to mass 
$M$ and $r_{t}$ by some later redshift $z < z_{\rm inf}$. We will denote the bound mass fraction $ \kappa(z,z_{\rm inf}) \equiv {M}/{M_{0}}$. 

Based on the results of \cite{Jiang2014}, we estimate the bound mass fraction as $\kappa(z,z_{\rm inf}) = \exp(-\frac{A\sqrt{\Delta_c}\Delta z}{\pi (1 + z)}),$ where A = 0.81, $\Delta_c$ is the mean overdensity of a halo relative to the critical density, assumed to be 200 in this work, $\Delta z$ is the difference between the redshift of infall $z_{inf}$ and the redshift of evaluation $z.$ In this parameterization of mass loss, we note that the bound mass fraction rapidly goes to zero as $\Delta z \rightarrow \infty$. Both earlier semi-analytic models \cite[e.g.][]{Taylor2005} and recent idealized simulations \cite{vdB2018} suggest subhalos may retain some mass for much longer times. Using an alternative parametrization $\kappa(0,z_{\rm inf}) = 0.2 + 0.8 \exp (-z_{\rm inf})$ based on the results of \cite{Taylor2005}, we find substructure boost factors 2--3 higher than those calculated using \cite{Jiang2014}, so tidal mass loss remains an uncertainty in our boost factor calculations.

Given an expression for the bound mass fraction, the tidal radius can be
calculated from the condition
$$\frac{M'(<r_t)}{M(<r)} = \kappa(z,z_{\rm inf})\,.$$
Note this implies that the tidal radius will depend on $z$, $z_{\rm inf}$, and the initial density profile at infall, parameterized by a concentration $c_{\rm inf}$ and/or 
a shape parameter $\alpha_{\rm inf}$.

To estimate the boost factor within a subhalo's tidal radius, we will use equation \ref{eq:boost_halo}, but with a density profile modified by tidal effects. 
The change to the profile is well-described by 
\begin{equation}
\rho_{\rm sub} = \frac{f_t}{1 + (r/r_t)^3}\rho_{\rm NFW},
\label{eq:hayashi}
\end{equation}
\cite{Hayashi2003}, where $f_t$ measures the change in the central density due to tidal heating, and $r_t$ is the tidal radius. For the purposes 
of our simple estimate here, we will ignore the change in central density, which is minor for moderate amounts of mass loss 
\cite{Hayashi2003,Drakos2017}, and take $f_t = 1$. Given this modified density profile, the boost from the subhalo is
\begin{equation}
B_{\rm sh} \equiv \frac{1}{\bar{\rho}_{\rm sub}^2 V}\int{\rho_{\rm sub}^2 dV}\,.
\end{equation}
While we calculate $B_{\rm sub}$ using the form in equation~\ref{eq:hayashi}, we note that since the stripping process produces an abrupt truncation at $r_t$, 
the modified boost factor is also approximately equal to the unstripped function $B_h(c^\prime)$, where $c^\prime \equiv r_t/r_s$ is a 
reduced concentration parameter incorporating the effects of tidal stripping. This reduced concentration factor will in turn depend 
on the initial concentration $c_{\rm inf}$, $z_{\rm inf}$, the redshift at which the subhalo merges, and $z$, the redshift at which the boost is measured.

\item{Density contrast relative to the main halo}

At the redshift $z_{\rm inf}$ when it is accreted into the host, the subhalo is assumed to be virialized, and thus should have an initial density of $\rho_{\rm vir}(z_{\rm inf}) = 200 \rho_c(z_{\rm inf})$
within its virial radius. By redshift $z$, tidal stripping will have increased this density by a factor $\kappa(z,z_{\rm inf}) (r_{\rm vir}/r_t)^3$. 
Thus the ratio of the subhalo density to the main halo density at $z$ is  
\begin{eqnarray}
\frac{\bar{\rho}_{\rm sub}(z,z_{\rm inf})}{\bar{\rho}_h} = \kappa(z,z_{\rm inf}) \left(\frac{r_{\rm vir}}{r_t}\right)^3\frac{\rho_{\rm vir}(z_{\rm inf})}{\rho_{\rm vir}(z)} = \kappa(z,z_{\rm inf}) \left(\frac{r_{\rm vir}}{r_t}\right)^3 \frac{\rho_{c}(z_{\rm inf})}{\rho_{c}(z)}\,.
\end{eqnarray}
Since the virial radius is determined implicitly from $\kappa(z,z_{\rm inf})$, it too will depend on $z$ and $z_{\rm inf}$, but not on subhalo mass, and thus the density contrast is also independent of subhalo mass in this model, but depends only on $z$ and $z_{\rm inf}$. 

 \item{Subhalo mass function and infall redshift distribution}

Finally, we need to specify the number of subhalos that merge to form the main halo as a function of mass and redshift. 
We will begin by calculating how much mass the main halo accretes at each redshift, and then assume that this accreted mass contains a distribution of subhalo masses proportional to the field halo mass function at that redshift.

The overall evolution of the main halo mass 
is described by its mass accretion history (MAH), M(z).
 
 Given a halo of mass $M(z_i)$ at redshift $z_i$, we will assume the mass $M(z)$ at any earlier epoch $z > z_i$ is given by the expression in \cite{Correa2015} 
 \begin{equation}
 M(z) = M(z_i) (1 + z - z_i)^{\tilde{\alpha}}e^{\tilde{\beta}(z-z_i)},
 \end{equation}
 where parameters $\tilde{\alpha}$ and $\tilde{\beta}$ depend on $M(z_i)$ and $z_i$ as 
 \begin{eqnarray}
\tilde{\alpha} &=& \left[\frac{1.686(2/\pi)^{1/2}}{D(z_i)^2}\frac{dD}{dz}|_{z=zi} + 1 \right]f(M(z_i)), \\
 \tilde{\beta} &=& - f(M(z_i)), \\
 D(z) &\propto& H(z) \int_z^{\infty} \frac{1+z'}{H(z')^3}dz', \\
 f(M(z_i)) &=& \left[\sigma^2(M(z_i)/q) - \sigma^2(M(z_i))\right]^{-1/2}, \\
 \sigma^2(R) &=& \frac{1}{2\pi^2}\int_0^{\infty}P(k)\hat{W}(k;R)k^2dk, \nonumber \\
 q &=& 4.137 \times z_f^{-0.9476}, \nonumber \\
 z_f &=& - 0.0064(\log_{10}M_0)^2 + 0.0237(\log_{10}M_0) + 1.8837.
 \end{eqnarray} 
 Here $H(z)$ is the Hubble parameter at a given redshift, $P(k)$ is the linear power spectrum, $\hat{W}(k;R)$ is the Fourier transform of a top-hat window function, and $D(z)$ is the linear growth factor normalized to unity at the present day. The parameters of the fit are valid for any cosmology.

The scatter in concentration discussed in section \ref{sec:2.3} is correlated with the MAH of individual halos, as discussed in \cite{Zhao09} -- more concentrated halos have older MAHs, while less concentrated systems built up their mass more recently. To model the effect of the scatter in individual MAHs on the subhalo boost factor, we use the simpler model of \cite{Wechsler2002}
\begin{equation}
M(z) = M_0 \exp(-2a_c z) = M_0 \exp\left(-\frac{8.2z}{c}\right)\,,
\end{equation}
which expresses the MAH in terms of $a_c$, the scale factor at the `formation epoch' of the halo, and $c$, the concentration of the halo at $z=0$. 
Given the concentration-mass relations of section~\ref{sec:2.2} and the scatter in concentration discussed in section~\ref{sec:2.3},
 we can calculate the corresponding mean and scatter in $a_c$ as a function of halo mass at $z=0$, and thus the effect on the boost factor.

In an interval $dz$, the increase in the main halo's mass $dM$ includes both bound and smooth material. Let $f_{bh}(M,z)$ be the fraction of the mass accreted 
in the form of bound halos with masses $m < M$. Due to tidal stripping, only a fraction $\kappa$ of this mass will survive to some later time 
as bound substructure. Thus, the total mass fraction in bound substructure at $z=0$ is given by 
\begin{equation}
X_{\rm sub}(z=0) = \frac{1}{M(0)}\int_0^{\infty} \kappa(0,z) f_{bh}[M(z),z]\frac{dM}{dz} dz,
\end{equation}
where $M(0)$ is the host halo mass at $z=0$. 
For redshifts different from zero, this may be generalized to 
\begin{equation}
X_{\rm sub}(z) = {1 \over{M (z)}}\int^\infty_{z } \kappa(z,z') f_{bh}[M(z'),(z')]\, {{dM}\over{dz'}} dz'
\end{equation}

Finally, to calculate the fraction of mass accreted as bound structure, $f_{bh}(M,z)$, we integrate the field halo mass function at $z$, from the mass of the 
smallest halo $M_{\rm lim}$ up to $M$ 
\begin{equation}
f_{bh}(M,z) = \int_{M_{\rm lim}}^{M}{\frac{M'}{\rho_m}\frac{dn}{dM'}(M',z)dM'},
\end{equation}
where $\frac{dn}{dM}(M,z)$ is the halo mass function (from \cite{Reed2007}) and $\rho_m$ is the mean matter density of the universe. 

We assume that the mass accreted as bound structure in a given redshift step consists of subhalos with a mass distribution that follows the (field) halo mass function $\frac{dn}{dM}(M,z)$ given by \cite{Reed2007}. This allows us to convert from a total mass fraction accreted at a given redshift to a set of individual subhalos accreted at that redshift. 

\end{enumerate}

Putting all these factors together, we can write an expression for the total boost, relative to the smooth halo boost, based on equation \ref{eqn:sub1} 
\begin{eqnarray}
\frac{B(V)}{B(V_0)}(z) &=&  \frac{B_t}{B_h}(z) = (1-X)  + (B_{\rm sub}/B_h)\\\nonumber
&=& (1-X)  + \sum_{i \, \in \, \rm subhaloes} {\left(\bar{\rho}_{\rm sub} (z,z_{{\rm inf},i})\over \bar{\rho}_h\right) \frac{dX_i}{dz}}  \frac{B_{{\rm sh},i}(z)}{B_h}\,.
\label{eqn:sub2}
\end{eqnarray}
Here the subhalos indexed with $i$ are the set of subhalos accreted in any one redshift step, summed over all redshift steps. While individual subhalos in a given step will vary in concentration, in practice, for a given $z$, $z_{\rm inf}$ and subhalo mass, we assume the mean (field halo) concentration $c(M,z_{\rm inf})$ to calculate the boost factor.

\subsection{Total substructure boost factor}
\label{sec:3.3}

\begin{figure}[t]
\centering
\includegraphics[width=0.9\textwidth]{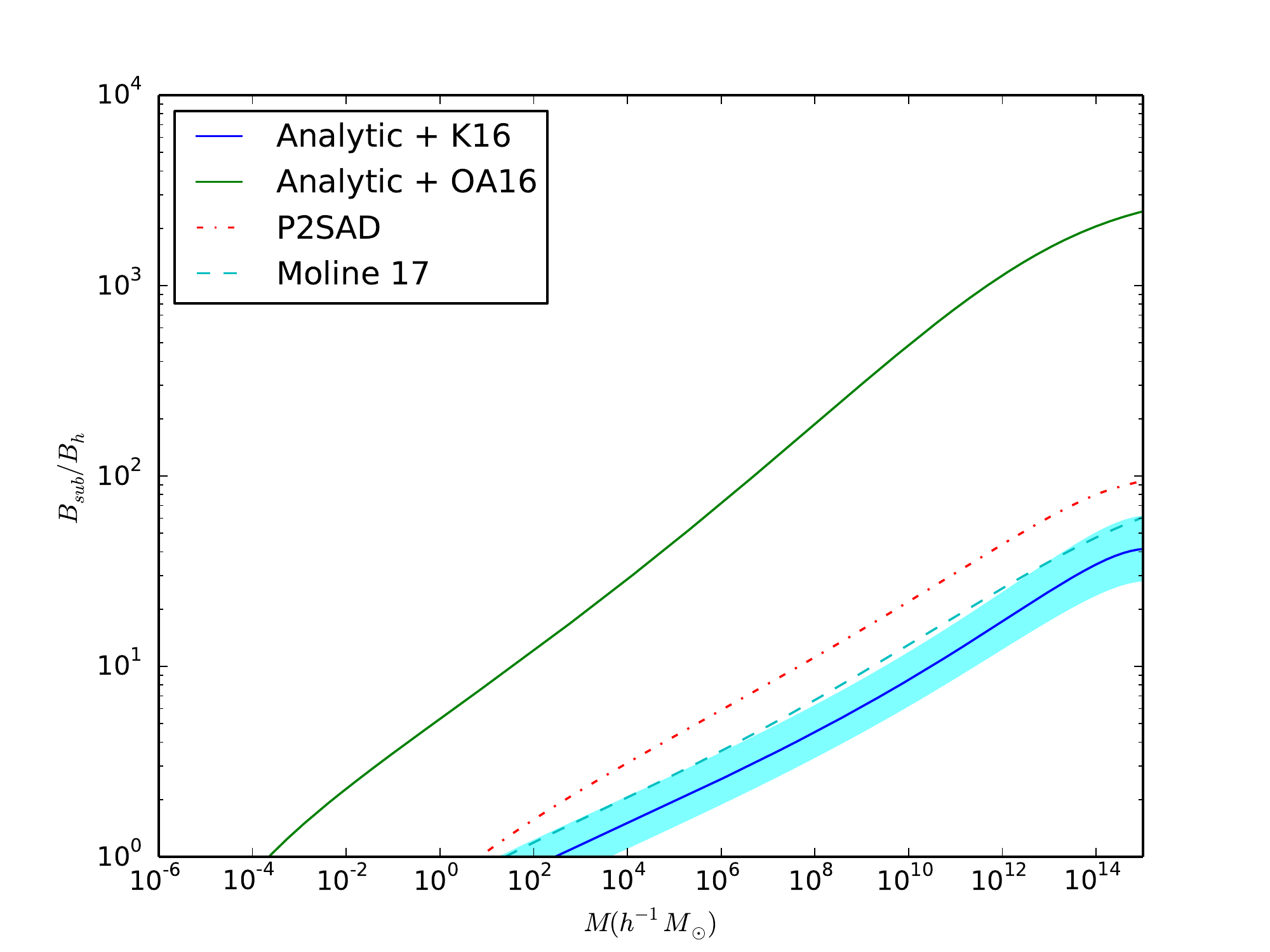}
\caption{The boost from substructure relative to that for a smooth halo, using two different prescriptions, $P^2SAD $(dot-dashed lines) and the analytic model (solid curves, for two different concentration-mass relations). Also plotted is the prediction from the analytic model of \cite{Moline2017} assuming a subhalo mass function of slope $\alpha = -2$ (long-dashed line). The cyan region illustrates the halo-to-halo scatter in boost factor around the analytic results with the lower concentration relation. The discrepancy between the green and blue curves reflects the assumption of an NFW profile, and will be alleviated if an Einasto profile is assumed instead (see figure 9 of OA16). Furthermore, an overall tidal dilution factor of $f_t \sim 0.01$ (equation 30 in \cite{P2SAD}) could explain the discrepancy between P2SAD and subhalo models.} 
\label{boost_substruct}
\end{figure}

The predictions for the substructure boost, relative to the smooth halo boost, are shown in figure \ref{boost_substruct}, as a function of halo mass, for $z=0$. In all cases, we assume an NFW profile and a lower mass limit $M_{\rm lim} = 10^{-6}M_\odot$. The solid (blue) curve is the analytic prediction, for the concentration-mass relation of K16. We see that despite the simplifying assumptions we have made, it is fairly similar to both the $P^2SAD$ predictions (dot-dashed lines) and to the recent estimate of \cite{Moline2017}. 
On the other hand, the uncertainties in the concentration-mass relation discussed in section~\ref{sec:2.2} lead to a large systematic uncertainty in the predictions. 
If we adopt then higher concentration-mass relation of OA16, 
the analytic predictions increase by more than one and a half orders of magnitude. This reflects the 0.75 dex disagreement in concentration at low mass seen in 
figure~\ref{fig:conc_range}, amplified by the power-law dependence of $B_h(c) \sim c^{2.5}$. 

%figure 6

The cyan band in figure~\ref{boost_substruct} shows the analytic model prediction for the halo-to-halo scatter in the substructure boost relative to the mean halo boost. The amplitude is approximately 0.17 dex, as compared to the scatter in the smooth halo boost factor due to concentration variations, which should be $\sim$0.45 dex, as 
discussed previously. Thus, if these two terms are uncorrelated we expect a scatter of $\sqrt{0.17^2 + 0.45^2} = 0.48$ dex (a factor of 3) in the total boost, including substructure. Since both terms correlate with formation epoch, however, it is possible that the total scatter is up to 0.62 dex (a factor of 4) if they act in the same direction, or 0.28 dex (a factor of $\sim$2) if they cancel.

\begin{figure}[t]
\centering
\includegraphics[width=0.9\textwidth]{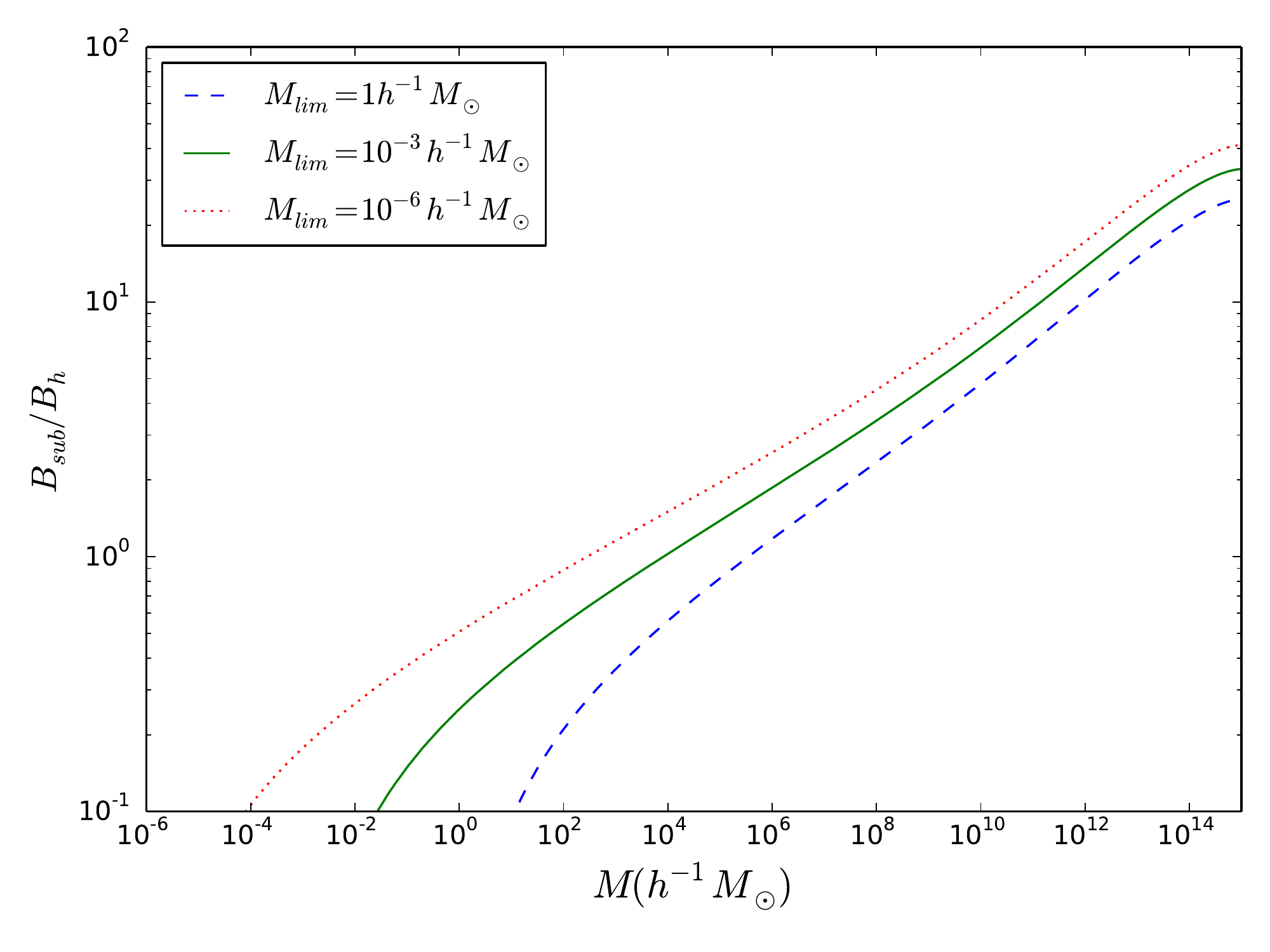}
\caption{The boost from substructure, relative to the smooth halo boost factor, as a function of halo mass, for three different values of the lower mass limit to CDM structure, $M_{\rm lim}$. In each case, we have assumed a NFW profile with the concentration-mass relation of K16.} 
\label{boost_mlim}
\end{figure}

In addition to the large uncertainty in the substructure boost factor due to the uncertainty in halo concentration at low mass, there is also a smaller uncertainty 
due to the possible range of $M_{\rm lim}$, the lower limit to the CDM mass function. This cutoff can vary from $10^{-3} M_\odot$ down to $10^{-12} M_\odot$ or less, depending 
on the mass and properties of the dark matter particle \cite[e.g.][]{Hofmann2001,Berezinsky2003,Green2004,Loeb2005,Bertschinger2006,Profumo2006,Bringmann2007,Bringmann2009}.
Figure~\ref{boost_mlim} shows how, as we vary $M_{\rm lim}$ upwards from
$10^{-6} h^{-1} M_\odot$ to $1 h^{-1} M_\odot$, the predicted boost drops by only 40\%. The uncertainty due to $M_{\rm lim}$, in turn, dominates 
uncertainties due to the cosmological parameters, which introduce smaller effects on the boost factor.

%figure 7

\subsection{Comparison with previous results}
\label{sec:3.4}

There have been many previous calculations of the substructure boost factor, starting with \cite{Ullio2002} and \cite{Taylor2003}. Here we compare our results to five recent versions of the calculation, by Bartels \&\ Ando \cite{Bartels2015}, Han et al.~\cite{Han2016}, Stref \&\ Lavalle \cite{Stref2017}, Moline et al.~\cite{Moline2017}, and Huetten et al.~\cite{Hutten2018}. Generally, these models all assume spherical halos and subhalos with NFW profiles (truncated at a tidal radius, in the case of the subhalos), and a power-law subhalo mass function. They make different assumptions, however, about the slope and amplitude of subhalo mass function, the rate of tidal mass loss, and the effective concentration parameter for subhalos. 

One controversial issue, as discussed in \cite{Bartels2015}, is what scale radius and scale density to assume for subhalo profiles. In some cases, models have started from the final subhalo mass at $z=0$, calculated a virial radius assuming the same density as for field galaxies at $z=0$ (i.e. $\sim 200\rho_c(0)$), and then used the present-day field halo concentration-mass relation to determine $r_{s}$ and $\rho_{s}$. This is approach clearly incorrect, for two separate reasons. First, the mean density is underestimated, since subhalos will be tidally stripped to smaller radii and larger densities than $z=0$ field halos, as discussed above. Second, with respect to the original virial radius of the subhalo when it fell in, the concentration assumed is too large on average, since the merger occurred at a higher redshift where mean concentrations at a given mass were smaller.

\cite{Bartels2015} develop an analytic model for the boost factor specifically to tackle this shortcoming of some previous studies. First, starting with an assumed present-day subhalo mass $m_0$ and tidal radius $r_t$ (or equivalently a concentration $c_t = r_s/r_t$), they evolve the subhalo backwards, assuming mass-loss rates from the model of  \cite{Jiang2014}, until the concentration-mass relation matches the field relation (taken from \cite{Correa2015}). This establishes a relationship between infall mass/redshift and the final values of $m_0$ and $c_t$. Applying this relationship to mass accretion histories and conditional subhalo accretion rates from \cite{Yang2011} and \cite{Correa2015} respectively, they then predict subhalo properties at $z=0$. 

Although their method predicts a final subhalo mass function slope and amplitude, they also consider predictions for a range of values of each. For slopes of the mass function $\alpha \equiv d\ln N/d\ln m = -1.9$ and $-2$, they predict substructure boost factors of roughly 5 and 30 for a cluster mass ($M_h = 10^{15}M_\odot$) main halo. Comparing to the results shown in figure \ref{boost_substruct}, we see that this is slightly below our analytic result using K16 concentrations, and a factor of 2 lower than the result from \cite{Moline2017}. The latter assumed $\alpha = -2$ and the same normalization for the subhalo mass function, but their concentration-mass relation is $\sim 0.3$ dex higher than that of \cite{Correa2015} at low masses, even for field halos, so this probably accounts for most of the difference. In our model, we take the subhalo (infall) mass function to be a scaled version of the field halo mass function, which gives it an effective slope $\alpha$ between $-1.95$ and $-2$ and a higher amplitude. This explains why our model  predicts a boost factor of $\sim 40$, even for the relatively low concentration-mass relation of K16.

\cite{Han2016} make a simple model for the subhalo population at $z=0$, and calibrate it using simulations of galaxy and cluster-mass halos. They assume that the initial (infall) mass function for subhalos, integrated of all redshifts, is a power-law, and fit the slope and amplitude assuming these also have a power-law dependence on the main halo mass. They calculate the boost factor assuming subhalo concentrations follow a present-day concentration-mass relation, but applied to their infall mass rather than their present-day mass, and show results for two different choices of relationship. Tidal stripping is assumed to scale as a power law of radius within the main halo, once again with parameters calibrated using simulations. Using the concentration-mass relation of \cite{2014MNRAS.441..378L}, they predict a boost factor of $\sim 40$ on cluster scales, which is very similar to our analytic results with  K16 concentrations. On the one hand, this concentration-mass relation is higher than K16; on the other hand the amplitude of their subhalo mass function is lower than ours, which compensates for this difference.

\cite{Stref2017} develop a model tuned to the particular case of the Milky Way, using dynamical constraints to fix the mass and structure of the main potential. Their model includes baryonic components, which produce additional tidal stripping due to enhanced global tides and rapid tidal `shocking' every time a subhalo passes through the galactic disk. They consider various possible slopes for the subhalo mass function, and also vary the lower mass limit for CDM structure. For concentrations, they assume the model of \cite{Sanchez-Conde2014}, which is similar to K16. While their focus is on the boost factor as a function of radius, we can compare to their predictions for the boost integrated out to the virial radius, which is $\sim3$ for $\alpha = -1.9$ or $\sim10$ for $\alpha = -2$. Given this is for a $M_h\sim10^{12}M_\odot$ halo, these results are a factor of two lower than ours; this may be because of the additional tidal heating effect of the baryonic components, although these tend to dominate only at small radii.

\cite{Moline2017} present a different approach to the problem of subhalo concentrations, measuring and fitting these directly from simulations, as a function of subhalo position within the main halo. The resulting concentration-mass relation is fairly high (similar to the curve labelled `Diemer mean' in figure \ref{boost_substruct}), even for subhalos close to the virial radius. For a subhalo mass function slope $\alpha = -2$, they predict a boost factor of $\sim 60$ on cluster scales, that is roughly twice the value we predict assuming K16 concentrations. The offset is presumably due to the concentration model, although our mass function slope is also slightly shallower than theirs.

Finally, \cite{Hutten2018} explore the sensitivity of the boost factor to different assumptions, as part of their calculation of the extragalactic gamma-ray background (EGB). Their default (conservative) model predicts a fairly low substructure boost factor ($\sim 5$) on cluster scales, but this is partly due to taking $\alpha = -1.9$, a fairly low normalization for the substructure mass function, and the low concentration-mass relation of \cite{Correa2015}. They also ignore the effect of mass loss on subhalo structure, which may reduce their effective subhalo densities. On the other hand, this paper nicely illustrates the relative uncertainties introduced by different components of the model. While the components of the model related to subhalos produce relatively little scatter in the EGB (which is dominated by the contribution from smaller halos, with lower boost factors), the overall range of predictions applied to a single halo is very roughly an order of magnitude.

We note that all these models make unrealistic assumptions about subhalos, such as abrupt tidal truncation, circular orbits, a steady-state distribution of satellite properties, pure power-law mass functions, etc.. In most cases, they are calibrated using simulations that only resolve subhalos down to mass scales of $\sim10^{6} M_\odot$, and extrapolate these results over many orders of magnitude to predict the total boost factor. In the end, however, the amplitude of the boost factor depends on the amount of material assumed to exist in the densest structures or regions. For halo-based models, this depends on the concentration-mass relation at low masses, the slope and normalization of the subhalo mass function, and the assumed efficiency of tidal stripping and disruption for the smallest objects. All of these model components are hard to determine precisely in current simulations, and require further numerical and analytic study.

Overall, boost factors calculated assuming traditional concentration-mass relations vary by about a factor of 5--10 below, or $\sim2$ above, our analytic results using K16 concentrations. This demonstrates a broad consistency between the detailed methods, with remaining uncertainties due to the unknown or poorly calibrated components of the models discussed above. The alternative concentration-mass model of OA16, or equivalently a model that assumes the central density of the first halos is conserved down to redshift zero even as they grow and merge, implies a much larger boost factor, as discussed previously. Thus, understanding the long-term evolution of the smallest halos is a crucial step in reducing uncertainties in the boost factor calculation.

 % section 4
\section{Baryonic sources and signal-to-noise in gamma-ray searches}
\label{sec:snr}

With increasing exposure, Fermi LAT has detected gamma-ray emission from a number of individual galaxies \cite[e.g.][]{Abdo2010,Ackermann2017}, and possibly also from some galaxy clusters \cite{Selig2015}. Most recently, \cite{Ackermann2017} found emission from the centre of M31 that could indicate dark matter annihilation, though it may also come from more mundane sources such as pulsars.
In general, for typical galaxies a stronger contaminant in dark matter annihilation searches are the contributions from the isotropic gamma-ray background (IGRB) \cite{Ackermann2015c}, ongoing star formation within the galaxy itself, and/or emission from an active galactic nucleus. While it should be possible to avoid galaxies with AGN and/or spatially resolve out this emission in the nearest galaxies, star formation is more ubiquitous and  spread out, and thus harder to avoid, while the contribution from the IGRB is unavoidable. Thus in this section, we will focus on star formation and the IGRB as the most important noise source in annihilation searches, at least for halos of group-scale or lower mass. 

Star-forming regions produce cosmic rays, whose interactions with molecular gas in turn lead to pion creation. The resulting decays produce gamma-ray emission at energies where it could mask dark matter annihilation. Earlier work by \cite{Abdo2010} and \cite{Ackermann2012} showed that the total gamma-ray luminosity seen by {\it Fermi LAT} above 100 MeV correlates closely with the star formation rate (SFR) for the nearby galaxies that have been detected individually. More recently, gamma-ray emission from star-forming galaxies has been considered in detail by a number of authors \cite[e.g.][]{Tamborra2014,Bechtol2017,Pfrommer2017}. For the purpose of this calculation, however, we will assume the simple empirical relationship of \cite{Abdo2010}:
\begin{equation}
L_{\rm SFR} = 7.4 \times {\rm (SFR)}^{1.4},
\label{eq:SFR}
\end{equation}
where $L$ is in units of $10^{41}$ photons s$^{-1}$,  and the SFR is in $M_{\odot} \, {\rm yr}^{-1}$.  
 
To estimate the signal-to-noise of the annihilation signal for an individual halo, we will estimate the stellar mass of its central galaxy, the typical star formation rate for 
an object with that stellar mass, and the resulting gamma-ray luminosity using the relation above. We will then compare this `noise' to the `signal', i.e.~the luminosity 
from annihilating dark matter in the halo.

We use the stellar-to-halo-mass-ratio (SHMR) relation of \cite{Behroozi2013} at $z=0$ to estimate the mean stellar mass of the central galaxy in a given halo.
This five-parameter function was derived by fitting the stellar mass functions and specific star formation rates of galaxies at redshifts $z=0$--8, 
as well as the cosmic star formation rate over this range. It may be expressed as 
\begin{equation}
\log_{10}(M_*(M_h)) = \log_{10}(\epsilon M_1) + f\left(\log_{10}\left[\frac{M_h}{M_1}\right]\right) - f(0),
\end{equation}
where 
$$f(x) \equiv - \log_{10}(10^{\alpha x} + 1) + \delta \frac{(\log_{10}(1 + \exp(x)))^{\gamma}}{1 + \exp(10^{-x})}\,.$$
 For $z=0$, the parameters, $\alpha$, $\delta$, $\gamma$, $\log_{10} \epsilon$, and $\log_{10} M_1$ have best-fit values 
-1.412, 3.508, 0.316, $-1.777$, and $11.514$ respectively. 

Given the stellar mass of the central galaxy, we then estimate its mean star formation rate, by assuming 
the galaxy lies on the `star formation main sequence'  \cite[e.g.][]{Santini2017}, such that 
\begin{equation}
\log_{10} {\rm SFR} = \alpha \log_{10}\left(\frac{M_*}{M_{9.7}}\right) + \beta,
\end{equation}
where $\alpha$, $\beta$, and $M_{9.7}$ are 1.04, 1.01 and $10^{9.7}M_{\odot}$ respectively,
and the SFR is in units of $M_{\odot} \, {\rm yr}^{-1}$. Given this rate, we can derive a gamma-ray luminosity due to star-formation from equation~\ref{eq:SFR} above.

The diffuse isotropic gamma-ray background (IGRB) is an all-sky gamma-ray emission from unresolved sources that remains after resolved sources, diffuse Galactic emission, the Cosmic Ray background, and the Solar contribution have been removed from the total all-sky background. Measurements of the IGRB using the LAT detector of the Fermi Gamma ray Space Telescope at energies from 100 MeV to 820 GeV from over fifty months of LAT data indicate an integrated intensity for energies above 100 MeV of $7.2 \times 10^{-6}$ photons ${\rm cm}^{-2} s^{-1} {\rm sr}^{-1}$ \cite{Ackermann2015c}. Using this intensity, the estimated luminosity from the IGRB from individual haloes is given by
\begin{equation}
L_{IGRB} = 7.2 \times 10^{-6} 4\pi R_{200}^2,
\end{equation} 
where $R_{200}$ is the virial radius of the halo.

The gamma-ray luminosity of the halo due to dark matter annihilation is given by the boost factor calculated previously, times the halo mass, mean density, and the
particle factors discussed in section~\ref{sec:2.1}:
\begin{equation}
L_h \equiv \frac{\langle \sigma v\rangle}{2m_{\chi}^2} B_{t}\ 200\rho_c(0) M_h N_{\gamma},
\end{equation}
where $B_{t}$ is the total boost from the halo, including the effects of substructure, $M_h$ is the mass of the halo and $N_{\gamma}$ is the number of photons above a given threshold, say 1 GeV, produced by a pair of WIMPs annihilating. 

Observing a source at distance $D$ for a time $\Delta t$ with a detector of effective area $A_{\rm det}$, 
the signal-to-noise ratio is then given as 
\begin{equation}
SNR = \frac{L_h A_{\rm det} \Delta t /4 \pi D^2}{\sqrt{(L_{\rm SFR} + L_h + L_{IGRB}})A_{\rm det} \Delta t /4 \pi D^2}\,.
\label{eq:SNR}
\end{equation}
where $A_{\rm det}$ is the Fermi LAT detector area, D is the estimated distance to a halo, and $\Delta t$ is the observation time.

\begin{figure}[t]
\centering
\includegraphics[width=0.9\textwidth]{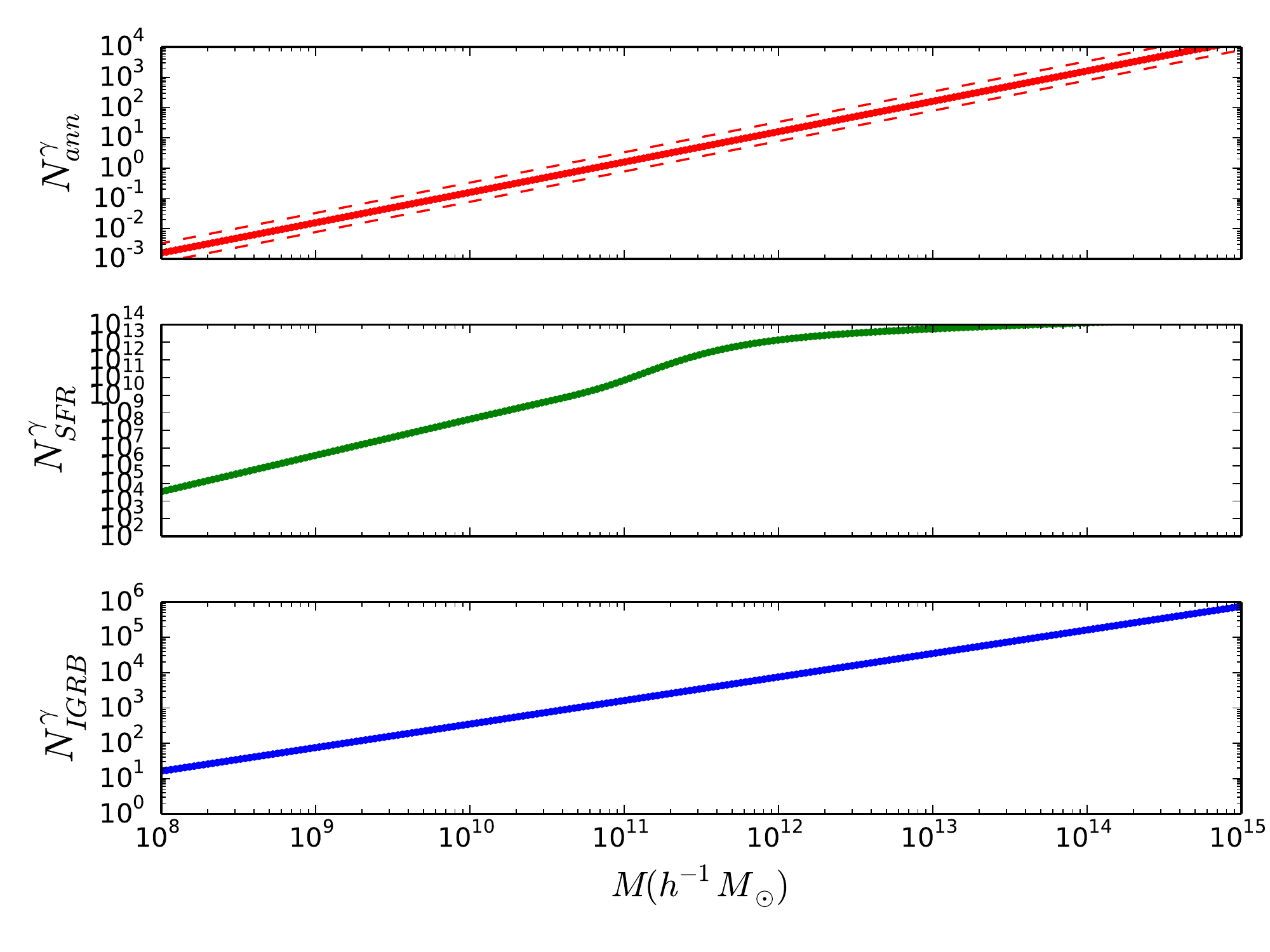}
\caption{Gamma rays produced by dark matter annihilation (top panel), star formation (middle panel), and from the IGRB (bottom panel), as a function of halo mass. The annihilation counts assume a velocity-averaged annihilation cross-section $\langle \sigma v\rangle = 3 \times 10^{-26}\,{\rm cm}^3\,{\rm s}^{-1}$, a WIMP mass of $m_{\chi} = 100$ GeV, 
$N_{\gamma} = 30$, and an observing time of $\Delta t = 8$ years, for the Fermi LAT (effective area $7200\, {\rm cm}^2$). The boost factor used assumes a NFW profile, 
the concentration relation of K16, and a mass limit of $M_{\rm lim} = 10^{-15}$. The star formation counts assume the luminosity-SFR relation from equation~\ref{eq:SFR}.} 
\label{ann_vs_sfr}
\end{figure}

As a concrete example, we will assume $\langle \sigma v\rangle = 3 \times 10^{-26}\,{\rm cm}^3\,{\rm s}^{-1},$ $m_{\chi} = 100$ GeV, $N_{\gamma} = 30$ and $\Delta t = 8$ years. The effective area for the Fermi LAT is $7200\, {\rm cm}^2$ \footnote{http://www.slac.stanford.edu/exp/glast/groups/canda/lat\_Performance.htm}. 
For these parameters, and for a source distance of $D = 10$ Mpc, figure~\ref{ann_vs_sfr} shows the expected counts from annihilation (top panel), from star formation (middle panel), and from the IGRB (bottom panel). The annihilation calculation assumes a NFW profile, the concentration relation of K16, and a mass limit of 
$M_{\rm lim} = 10^{-15} M_\odot$.
Note the predicted signal is small; for galaxy mass halos we expect only $\sim$10 photons/year from annihilation, given the parameters 
we have chosen. (We also note that the scatter in the annihilation counts is only a factor of $\sim$2.5, suggesting the effects of concentration 
and substructure variations are slightly anti-correlated -- see the discussion in section~\ref{sec:3.3}.)

%figure 8

We can consider the SNR in two cases, either that we observe different halos at a fixed distance, or that for each halo mass, we find and observe the nearest system.
The average distance to the nearest halo of mass $M$ in a cosmological volume will be related to the number density of halos of that mass 
by $ D = n(M)^{-1/3}$. Thus, given a halo mass function $n(M)$,  we can calculate $D(M)$ and use this in equation~\ref{eq:SNR} to get the SNR for a typical 
closest source, as a function of halo mass. 

\begin{figure}[t]
\centering
\includegraphics[width=0.9\textwidth]{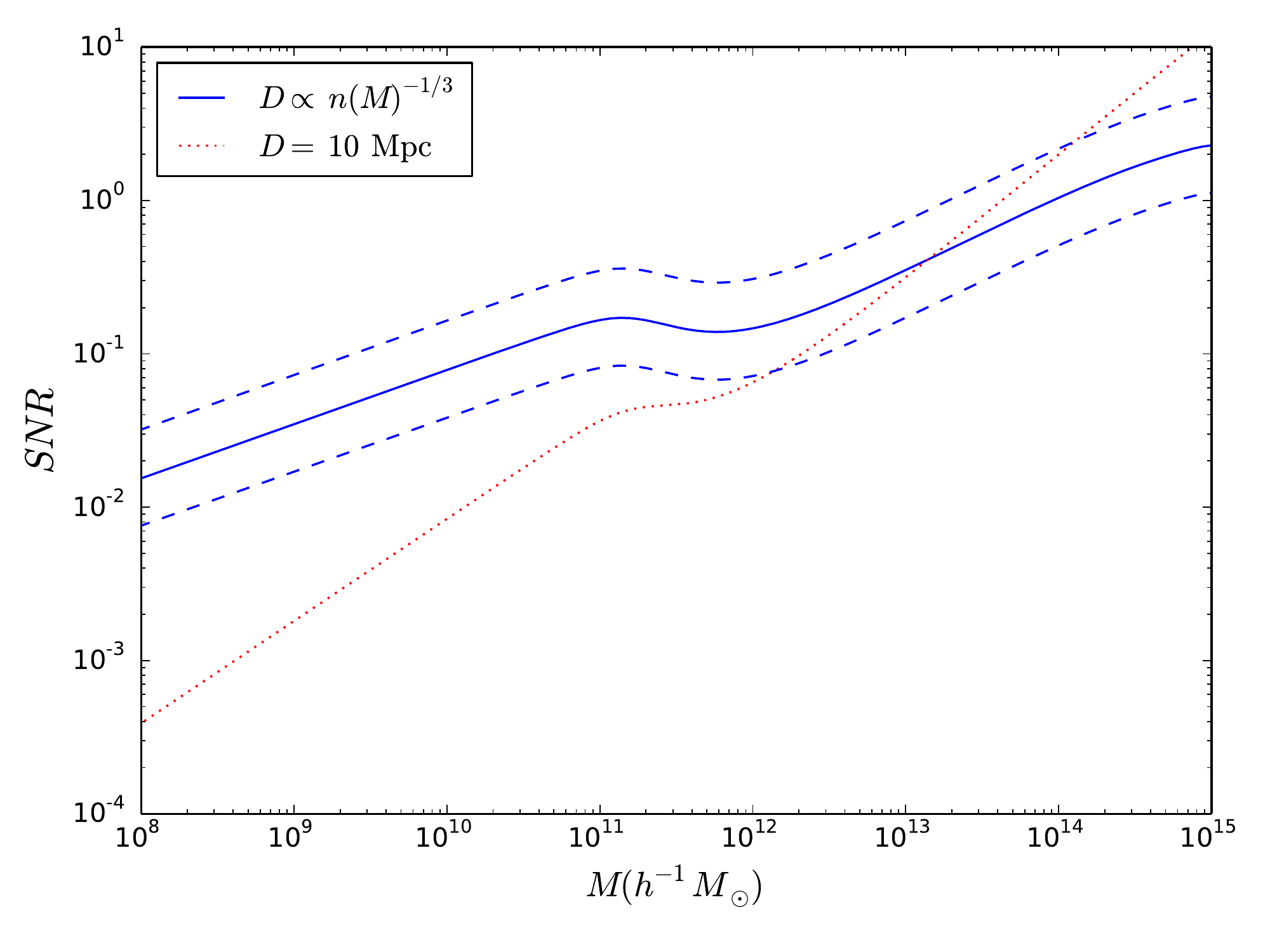}
\caption{The SNR of the annihilation signal from a single halo in our fiducial calculation, as a function of halo mass. The solid (blue) curve assumes we observe the nearest halo 
of that mass, at a distance scaling as $ D = n(M)^{-1/3}$. The dotted (red) curve assumes we observe halos at a fixed distance of 10 Mpc, independent of mass. 
The dashed lines show the 68\%\ halo-to-halo scatter, due primarily to differences in concentration.} 
\label{snr}
\end{figure}

Figure~\ref{snr} shows how SNR varies with halo mass, both at a fixed distance (dotted curve), and for the nearest halo of a given mass (solid curve, with dashed curves indicating halo-to-halo scatter). In the former case, for our chosen parameters and a distance of 10 Mpc, we see that only group-scale or larger systems are detectable individually at SNR $> 1$, and only massive clusters are detectable at SNR $\gtrsim 3$. If we consider the nearest halos, on 
the one hand, only clusters are detectable at SNR $> 1$; on the other hand the trend with mass is much shallower. Stacked samples of dwarf galaxies,
corresponding to halos in the mass range $10^9$--$10^{11}$, have so far produced the most stringent limits on the annihilation cross-section. 
This plot suggests that stacking the nearest groups or clusters with masses $M_h = 10^{13}$--$10^{14}M_\odot$ might achieve comparable limits.  

%figure 9

Finally, the dashed (blue) curves on figure~\ref{snr} show the importance of halo-to-halo scatter; considering the effect on dark matter structure alone, there should be 
a factor of 2 or more scatter in the SNR for individual halos. Since the primary origin of this scatter is the variation in halo concentration, independent estimates of this 
quantity could help select more promising objects for stacking. Variations in SFR will increase this variation further. For fixed dark matter properties, 
targeting passive galaxies in concentrated halos may yield SNRs many times those of the median relation.

%section 5
\section{Conclusion}
\label{sec:discuss}

Gamma-ray emission suggestive of dark matter annihilation has been identified in Fermi LAT maps of the centre of our galaxy \cite{Goodenough2009}, and more recently in maps of M31 \cite{Ackermann2017}. While these are expected to be the strongest sources of this kind, in both cases the emission is also consistent with other hypotheses, such as emission from pulsars. With increasing exposure, other nearby galaxies, groups and clusters are becoming promising targets for further analysis. Constraints based on the joint analysis of the Milky Way dwarfs are currently stronger than those expected from individual extragalactic halos \cite{Charles2016}, but stacking and halo-to-halo scatter could make the latter more competitive. As an example, \cite{Lisanti2017} recently stacked eight years of {\it Fermi} LAT data around a sample of groups and clusters out to $z=0.003$ (i.e.~D$\,\sim\,$130 Mpc). Even assuming a fairly modest (total) boost factor of $B\sim$5, they obtain constraints close to those obtained from dwarf satellites, particularly at low energies. Most of the improvement in their limits comes from the strongest 8--10 sources (although this depends slightly on energy range), so given the 
large halo-to-halo scatter expected, joint constraints on the strongest sources considered individually might be even more sensitive than the stacked result. Either way, annihilation searches in extragalactic halos have an advantage over those centred on dwarf satellites, that the underlying mass distribution is better 
known from independent dynamical measurements, and the interpretation of any limits or detection is thus less controversial.

To calculate the expected signal from individual extra-galactic halos, one needs to estimate the boost factors both for smooth halos and for substructure within halos. Previous calculations have generally considered the mean boost at a given mass. Here we develop an analytic formalism that includes halo-to-halo scatter, parameterized as a variation in concentration due to a variation in formation epoch. We include in the model the latest information about the halo density profile and about halo concentration as a function of mass and redshift. For our adopted (Planck 2015) cosmology, whereas other recent work predicts that concentration reaches a maximum value of $c\sim40$--70 around mass scales of $M_{h}\sim10^{-7}$--$10^{-9}$, the model of OA16
%\cite{Okoli2016} 
suggests a much higher value of $c\sim240$. We argue that this higher value is plausible, given the central densities measured in simulations of microhalo formation \cite{Ishiyama2014}. If these densities are conserved in field halos down to redshift zero, or if they survive as the first halos are incorporated into larger systems as substructure, then the expected boost will be 4--$6^{2.5} \sim 30$--90 times larger than previously predicted. This uncertainty dominates  others in the calculation, including the uncertainty in the form of the density profile, and in the limiting smallest mass scale for CDM structure. 

Modulo the uncertainty in the concentration of the smallest structures at low redshift, our analytic estimate for the boost factor is in good agreement both with other recent estimates \cite[e.g.][]{Moline2017}, and with the $P^2SAD$ model of \cite{P2SAD}, which is based on a completely different approach. We conclude that further work should focus not on  the hierarchical assembly process, which seems reasonably well understood (at least for the purposes of calculating the boost factor), but on the question of halo density at low mass and how it evolves from high redshift to low redshift. Unfortunately,  because microhalos form on mass scales that go non-linear long before $z=0$, this problem is not amenable to direct simulation, but requires some combination of numerical, analytic and/or semi-analytic work. Studies of microhalo formation also find that the initial density profile for the very first generation of halos have a steeper central cusp \cite{Ishiyama2014}; this probably has a smaller effect on the boost factor, but also requires further investigation. 

Given a conservative estimate for the boost factor, we calculate the signal-to-noise ratio for the detection of individual halos in the nearby Universe, assuming star formation is the main source of contaminating gamma-ray emission. At a given distance, the most massive halos have the highest SNR, but given low-mass halos are more common, we recover the result that the strongest constraints should come from stacked analyses at the (local) dwarf galaxy scale, or from individual analyses of the most massive clusters at greater distances. These very generic predictions will be modulated by the actual distribution of objects with mass and distance, but also by their overall star formation rate. For a low boost factor,  only very 
massive halos produce $SNR>3$, but for a high value much of the halo mass range could produce detections, given a WIMP mass of 100 Gev and a thermal cross-section. The non-detection of a clear signal from local halos would constrain the WIMP cross-section and mass correspondingly. Given the continuing search for the annihilation signal with Fermi LAT, with future satellite missions such as GAMMA-400\footnote{http://gamma400.lebedev.ru/indexeng.html}, or, at higher energies, with ground-based ACTs such as the Cherenkov Telescope Array \cite{Acharya2013}, it is important to resolve the remaining uncertainties in theoretical predictions of the annihilation rate. In particular, further work is needed to determine how the central densities of the smallest halos evolve from high redshift down to low redshift, in order to obtain definitive values for the 
substructure boost factor. 

\section*{Acknowledgements}
The authors acknowledge helpful discussions with J.~Zavala. CO and NA are supported by the University of Waterloo and the Perimeter Institute for Theoretical Physics. Research at Perimeter Institute is supported by the Government of Canada through Industry Canada and by the Province of Ontario through the Ministry of Research \&\ Innovation. JET acknowledges support from the Natural Science and Engineering Research Council of Canada, through a Discovery Grant.

\bibliographystyle{unsrtnat}
\bibliography{full_draft_4}

\begin{thebibliography}{130}
\providecommand{\natexlab}[1]{#1}
\providecommand{\url}[1]{\texttt{#1}}
\expandafter\ifx\csname urlstyle\endcsname\relax
  \providecommand{\doi}[1]{doi: #1}\else
  \providecommand{\doi}{doi: \begingroup \urlstyle{rm}\Url}\fi

\bibitem[{Planck Collaboration} and {Ade et al.}(2016)]{Planck2015}
{Planck Collaboration} and P.~A.~R. {Ade et al.}
\newblock {Planck 2015 results. XIII. Cosmological parameters}.
\newblock \emph{\aap}, 594:\penalty0 A13, September 2016.
\newblock \doi{10.1051/0004-6361/201525830}.

\bibitem[{Alam, et al.}(2017)]{Alam2017}
S.~{Alam, et al.}
\newblock {The clustering of galaxies in the completed SDSS-III Baryon
  Oscillation Spectroscopic Survey: cosmological analysis of the DR12 galaxy
  sample}.
\newblock \emph{\mnras}, 470:\penalty0 2617--2652, September 2017.
\newblock \doi{10.1093/mnras/stx721}.

\bibitem[{Joudaki, et al.}(2017)]{Joudaki2017}
S.~{Joudaki, et al.}
\newblock {KiDS-450 + 2dFLenS: Cosmological parameter constraints from weak
  gravitational lensing tomography and overlapping redshift-space galaxy
  clustering}.
\newblock \emph{ArXiv e-prints}, July 2017.

\bibitem[{Umetsu} et~al.(2015){Umetsu}, {Sereno}, {Medezinski}, {Nonino},
  {Mroczkowski}, {Diego}, {Ettori}, {Okabe}, {Broadhurst}, and
  {Lemze}]{Umetsu2015}
K.~{Umetsu}, M.~{Sereno}, E.~{Medezinski}, M.~{Nonino}, T.~{Mroczkowski}, J.~M.
  {Diego}, S.~{Ettori}, N.~{Okabe}, T.~{Broadhurst}, and D.~{Lemze}.
\newblock {Three-dimensional Multi-probe Analysis of the Galaxy Cluster A1689}.
\newblock \emph{\apj}, 806:\penalty0 207, June 2015.
\newblock \doi{10.1088/0004-637X/806/2/207}.

\bibitem[{Biviano} et~al.(2017){Biviano}, {Popesso}, {Dietrich}, {Zhang},
  {Erfanianfar}, {Romaniello}, and {Sartoris}]{Biviano2017}
A.~{Biviano}, P.~{Popesso}, J.~P. {Dietrich}, Y.-Y. {Zhang}, G.~{Erfanianfar},
  M.~{Romaniello}, and B.~{Sartoris}.
\newblock {Abell 315: reconciling cluster mass estimates from kinematics,
  X-ray, and lensing}.
\newblock \emph{\aap}, 602:\penalty0 A20, June 2017.
\newblock \doi{10.1051/0004-6361/201629471}.

\bibitem[{Lagattuta, et al.}(2017)]{Lagattuta2017}
D.~J. {Lagattuta, et al.}
\newblock {Lens modelling Abell 370: crowning the final frontier field with
  MUSE}.
\newblock \emph{\mnras}, 469:\penalty0 3946--3964, August 2017.
\newblock \doi{10.1093/mnras/stx1079}.

\bibitem[{Leauthaud, et al.}(2012)]{Leauthaud2012}
A.~{Leauthaud, et al.}
\newblock {New Constraints on the Evolution of the Stellar-to-dark Matter
  Connection: A Combined Analysis of Galaxy-Galaxy Lensing, Clustering, and
  Stellar Mass Functions from z = 0.2 to z =1}.
\newblock \emph{\apj}, 744:\penalty0 159, January 2012.
\newblock \doi{10.1088/0004-637X/744/2/159}.

\bibitem[{Velander, et al.}(2014)]{Velander2014}
M.~{Velander, et al.}
\newblock {CFHTLenS: the relation between galaxy dark matter haloes and baryons
  from weak gravitational lensing}.
\newblock \emph{\mnras}, 437:\penalty0 2111--2136, January 2014.
\newblock \doi{10.1093/mnras/stt2013}.

\bibitem[{Hudson, et al.}(2015)]{Hudson2015}
M.~J. {Hudson, et al.}
\newblock {CFHTLenS: co-evolution of galaxies and their dark matter haloes}.
\newblock \emph{\mnras}, 447:\penalty0 298--314, February 2015.
\newblock \doi{10.1093/mnras/stu2367}.

\bibitem[{van Uitert, et al.}(2016)]{vanUitert2016}
E.~{van Uitert, et al.}
\newblock {The stellar-to-halo mass relation of GAMA galaxies from 100
  deg$^{2}$ of KiDS weak lensing data}.
\newblock \emph{\mnras}, 459:\penalty0 3251--3270, July 2016.
\newblock \doi{10.1093/mnras/stw747}.

\bibitem[{More} et~al.(2011){More}, {van den Bosch}, {Cacciato}, {Skibba},
  {Mo}, and {Yang}]{More2011}
S.~{More}, F.~C. {van den Bosch}, M.~{Cacciato}, R.~{Skibba}, H.~J. {Mo}, and
  X.~{Yang}.
\newblock {Satellite kinematics - III. Halo masses of central galaxies in
  SDSS}.
\newblock \emph{\mnras}, 410:\penalty0 210--226, January 2011.
\newblock \doi{10.1111/j.1365-2966.2010.17436.x}.

\bibitem[{Ouellette, et al.}(2017)]{Ouellette2017}
N.~N.-Q. {Ouellette, et al.}
\newblock {The Spectroscopy and H-band Imaging of Virgo Cluster Galaxies
  (SHIVir) Survey: Scaling Relations and the Stellar-to-total Mass Relation}.
\newblock \emph{\apj}, 843:\penalty0 74, July 2017.
\newblock \doi{10.3847/1538-4357/aa74b1}.

\bibitem[{McConnachie}(2012)]{McConnachie2012}
A.~W. {McConnachie}.
\newblock {The Observed Properties of Dwarf Galaxies in and around the Local
  Group}.
\newblock \emph{\aj}, 144:\penalty0 4, July 2012.
\newblock \doi{10.1088/0004-6256/144/1/4}.

\bibitem[{Feng}(2010)]{Feng2010}
J.~L. {Feng}.
\newblock {Dark Matter Candidates from Particle Physics and Methods of
  Detection}.
\newblock \emph{\araa}, 48:\penalty0 495--545, September 2010.
\newblock \doi{10.1146/annurev-astro-082708-101659}.

\bibitem[{Gaskins}(2016)]{Gaskins2016}
J.~M. {Gaskins}.
\newblock {A review of indirect searches for particle dark matter}.
\newblock \emph{Contemporary Physics}, 57:\penalty0 496--525, October 2016.
\newblock \doi{10.1080/00107514.2016.1175160}.

\bibitem[{Cushman, et al.}(2013)]{Snowmass2013}
P.~{Cushman, et al.}
\newblock {Snowmass CF1 Summary: WIMP Dark Matter Direct Detection}.
\newblock \emph{ArXiv e-prints}, October 2013.

\bibitem[{Marrod{\'a}n Undagoitia} and {Rauch}(2016)]{Marrodan2016}
T.~{Marrod{\'a}n Undagoitia} and L.~{Rauch}.
\newblock {Dark matter direct-detection experiments}.
\newblock \emph{Journal of Physics G Nuclear Physics}, 43\penalty0
  (1):\penalty0 013001, January 2016.
\newblock \doi{10.1088/0954-3899/43/1/013001}.

\bibitem[{MAGIC Collaboration}(2016)]{Ahnen2016}
{MAGIC Collaboration}.
\newblock {Limits to dark matter annihilation cross-section from a combined
  analysis of MAGIC and Fermi-LAT observations of dwarf satellite galaxies}.
\newblock \emph{\jcap}, 2:\penalty0 039, February 2016.
\newblock \doi{10.1088/1475-7516/2016/02/039}.

\bibitem[{Charles, et al.}(2016)]{Charles2016}
E.~{Charles, et al.}
\newblock {Sensitivity projections for dark matter searches with the Fermi
  large area telescope}.
\newblock \emph{\physrep}, 636:\penalty0 1--46, June 2016.
\newblock \doi{10.1016/j.physrep.2016.05.001}.

\bibitem[{Archer, et al.}(2014)]{Archer2014}
A.~{Archer, et al.}
\newblock {Very-high Energy Observations of the Galactic Center Region by
  VERITAS in 2010-2012}.
\newblock \emph{\apj}, 790:\penalty0 149, August 2014.
\newblock \doi{10.1088/0004-637X/790/2/149}.

\bibitem[{Abdallah, et al.}(2016)]{Abdallah2016}
H.~{Abdallah, et al.}
\newblock {Search for Dark Matter Annihilations towards the Inner Galactic Halo
  from 10 Years of Observations with H.E.S.S.}
\newblock \emph{Physical Review Letters}, 117\penalty0 (11):\penalty0 111301,
  September 2016.
\newblock \doi{10.1103/PhysRevLett.117.111301}.

\bibitem[{Ajello, et al.}(2016)]{Ajello2016}
M.~{Ajello, et al.}
\newblock {Fermi-LAT Observations of High-Energy Gamma-Ray Emission toward the
  Galactic Center}.
\newblock \emph{\apj}, 819:\penalty0 44, March 2016.
\newblock \doi{10.3847/0004-637X/819/1/44}.

\bibitem[{Gondolo}(1994)]{Gondolo1994}
P.~{Gondolo}.
\newblock {Dark matter annihilations in the Large Magellanic Cloud}.
\newblock \emph{Nuclear Physics B Proceedings Supplements}, 35:\penalty0
  148--149, May 1994.
\newblock \doi{10.1016/0920-5632(94)90234-8}.

\bibitem[{Baltz} et~al.(2000){Baltz}, {Briot}, {Salati}, {Taillet}, and
  {Silk}]{Baltz2000}
E.~A. {Baltz}, C.~{Briot}, P.~{Salati}, R.~{Taillet}, and J.~{Silk}.
\newblock {Detection of neutralino annihilation photons from external
  galaxies}.
\newblock \emph{\prd}, 61\penalty0 (2):\penalty0 023514, January 2000.
\newblock \doi{10.1103/PhysRevD.61.023514}.

\bibitem[{Fornengo} et~al.(2004){Fornengo}, {Pieri}, and
  {Scopel}]{Fornengo2004}
N.~{Fornengo}, L.~{Pieri}, and S.~{Scopel}.
\newblock {Neutralino annihilation into {$\gamma$} rays in the Milky Way and in
  external galaxies}.
\newblock \emph{\prd}, 70\penalty0 (10):\penalty0 103529, November 2004.
\newblock \doi{10.1103/PhysRevD.70.103529}.

\bibitem[{Tasitsiomi} et~al.(2004){Tasitsiomi}, {Gaskins}, and
  {Olinto}]{Tasitsiomi2004}
A.~{Tasitsiomi}, J.~{Gaskins}, and A.~V. {Olinto}.
\newblock {Gamma-ray and synchrotron emission from neutralino annihilation in
  the Large Magellanic Cloud}.
\newblock \emph{Astroparticle Physics}, 21:\penalty0 637--650, September 2004.
\newblock \doi{10.1016/j.astropartphys.2004.04.004}.

\bibitem[{Mack} et~al.(2008){Mack}, {Jacques}, {Beacom}, {Bell}, and
  {Y{\"u}ksel}]{Mack2008}
G.~D. {Mack}, T.~D. {Jacques}, J.~F. {Beacom}, N.~F. {Bell}, and
  H.~{Y{\"u}ksel}.
\newblock {Conservative constraints on dark matter annihilation into gamma
  rays}.
\newblock \emph{\prd}, 78\penalty0 (6):\penalty0 063542, September 2008.
\newblock \doi{10.1103/PhysRevD.78.063542}.

\bibitem[{Saxena} et~al.(2011){Saxena}, {Summa}, {Els{\"a}sser}, {R{\"u}ger},
  and {Mannheim}]{Saxena2011}
S.~{Saxena}, A.~{Summa}, D.~{Els{\"a}sser}, M.~{R{\"u}ger}, and K.~{Mannheim}.
\newblock {Constraints on dark matter annihilation from M87. Signatures of
  prompt and inverse-Compton gamma rays}.
\newblock \emph{European Physical Journal C}, 71:\penalty0 1815, November 2011.
\newblock \doi{10.1140/epjc/s10052-011-1815-y}.

\bibitem[{Buckley} et~al.(2015){Buckley}, {Charles}, {Gaskins}, {Brooks},
  {Drlica-Wagner}, {Martin}, and {Zhao}]{Buckley2015}
M.~R. {Buckley}, E.~{Charles}, J.~M. {Gaskins}, A.~M. {Brooks},
  A.~{Drlica-Wagner}, P.~{Martin}, and G.~{Zhao}.
\newblock {Search for gamma-ray emission from dark matter annihilation in the
  large magellanic cloud with the fermi large area telescope}.
\newblock \emph{\prd}, 91\penalty0 (10):\penalty0 102001, May 2015.
\newblock \doi{10.1103/PhysRevD.91.102001}.

\bibitem[{Caputo, et al.}(2016)]{Caputo2016}
R.~{Caputo, et al.}
\newblock {Search for gamma-ray emission from dark matter annihilation in the
  Small Magellanic Cloud with the Fermi Large Area Telescope}.
\newblock \emph{\prd}, 93\penalty0 (6):\penalty0 062004, March 2016.
\newblock \doi{10.1103/PhysRevD.93.062004}.

\bibitem[{Li} et~al.(2016){Li}, {Huang}, {Yuan}, and {Xu}]{LiHuang2016}
Z.~{Li}, X.~{Huang}, Q.~{Yuan}, and Y.~{Xu}.
\newblock {Constraints on the dark matter annihilation from Fermi-LAT
  observation of M31}.
\newblock \emph{\jcap}, 12:\penalty0 028, December 2016.
\newblock \doi{10.1088/1475-7516/2016/12/028}.

\bibitem[{Geringer-Sameth} and {Koushiappas}(2011)]{GeringerSameth2011}
A.~{Geringer-Sameth} and S.~M. {Koushiappas}.
\newblock {Exclusion of Canonical Weakly Interacting Massive Particles by Joint
  Analysis of Milky Way Dwarf Galaxies with Data from the Fermi Gamma-Ray Space
  Telescope}.
\newblock \emph{Physical Review Letters}, 107\penalty0 (24):\penalty0 241303,
  December 2011.
\newblock \doi{10.1103/PhysRevLett.107.241303}.

\bibitem[{Ackermann, et al.}(2011)]{Ackermann2011}
M.~{Ackermann, et al.}
\newblock {Constraining Dark Matter Models from a Combined Analysis of Milky
  Way Satellites with the Fermi Large Area Telescope}.
\newblock \emph{Physical Review Letters}, 107\penalty0 (24):\penalty0 241302,
  December 2011.
\newblock \doi{10.1103/PhysRevLett.107.241302}.

\bibitem[{Ackermann, et al.}(2014)]{Ackermann2014}
M.~{Ackermann, et al.}
\newblock {Dark matter constraints from observations of 25 Milky{\^A} Way
  satellite galaxies with the Fermi Large Area Telescope}.
\newblock \emph{\prd}, 89\penalty0 (4):\penalty0 042001, February 2014.
\newblock \doi{10.1103/PhysRevD.89.042001}.

\bibitem[{Ackermann, et al.}(2015{\natexlab{a}})]{Ackermann2015}
M.~{Ackermann, et al.}
\newblock {Searching for Dark Matter Annihilation from Milky Way Dwarf
  Spheroidal Galaxies with Six Years of Fermi Large Area Telescope Data}.
\newblock \emph{Physical Review Letters}, 115\penalty0 (23):\penalty0 231301,
  December 2015{\natexlab{a}}.
\newblock \doi{10.1103/PhysRevLett.115.231301}.

\bibitem[{Albert et al.}(2017)]{Albert2017}
A.~{Albert et al.}
\newblock {Dark Matter Limits From Dwarf Spheroidal Galaxies with The HAWC
  Gamma-Ray Observatory}.
\newblock \emph{ArXiv e-prints}, June 2017.

\bibitem[{Ackermann, et al.}(2010)]{Ackermann2010}
M.~{Ackermann, et al.}
\newblock {Constraints on dark matter annihilation in clusters of galaxies with
  the Fermi large area telescope}.
\newblock \emph{\jcap}, 5:\penalty0 025, May 2010.
\newblock \doi{10.1088/1475-7516/2010/05/025}.

\bibitem[{Arlen, et al.}(2012)]{Arlen2012}
T.~{Arlen, et al.}
\newblock {Constraints on Cosmic Rays, Magnetic Fields, and Dark Matter from
  Gamma-Ray Observations of the Coma Cluster of Galaxies with VERITAS and
  Fermi}.
\newblock \emph{\apj}, 757:\penalty0 123, October 2012.
\newblock \doi{10.1088/0004-637X/757/2/123}.

\bibitem[{Huang} et~al.(2012){Huang}, {Vertongen}, and {Weniger}]{Huang2012}
X.~{Huang}, G.~{Vertongen}, and C.~{Weniger}.
\newblock {Probing dark matter decay and annihilation with Fermi LAT
  observations of nearby galaxy clusters}.
\newblock \emph{\jcap}, 1:\penalty0 042, January 2012.
\newblock \doi{10.1088/1475-7516/2012/01/042}.

\bibitem[{Prokhorov}(2014)]{Prokhorov2014}
D.~A. {Prokhorov}.
\newblock {An analysis of Fermi-LAT observations of the outskirts of the Coma
  cluster of galaxies}.
\newblock \emph{\mnras}, 441:\penalty0 2309--2315, July 2014.
\newblock \doi{10.1093/mnras/stu707}.

\bibitem[{Ackermann, et al.}(2015{\natexlab{b}})]{Ackermann2015b}
M.~{Ackermann, et al.}
\newblock {Search for Extended Gamma-Ray Emission from the Virgo Galaxy Cluster
  with FERMI-LAT}.
\newblock \emph{\apj}, 812:\penalty0 159, October 2015{\natexlab{b}}.
\newblock \doi{10.1088/0004-637X/812/2/159}.

\bibitem[{Ackermann, et al.}(2016)]{Ackermann2016b}
M.~{Ackermann, et al.}
\newblock {Search for Gamma-Ray Emission from the Coma Cluster with Six Years
  of Fermi-LAT Data}.
\newblock \emph{\apj}, 819:\penalty0 149, March 2016.
\newblock \doi{10.3847/0004-637X/819/2/149}.

\bibitem[{Liang, et al.}(2016)]{Liang2016}
Y.-F. {Liang, et al.}
\newblock {Search for a gamma-ray line feature from a group of nearby galaxy
  clusters with Fermi LAT Pass 8 data}.
\newblock \emph{\prd}, 93\penalty0 (10):\penalty0 103525, May 2016.
\newblock \doi{10.1103/PhysRevD.93.103525}.

\bibitem[{Anderson} et~al.(2016){Anderson}, {Zimmer}, {Conrad}, {Gustafsson},
  {S{\'a}nchez-Conde}, and {Caputo}]{Anderson2016}
B.~{Anderson}, S.~{Zimmer}, J.~{Conrad}, M.~{Gustafsson},
  M.~{S{\'a}nchez-Conde}, and R.~{Caputo}.
\newblock {Search for gamma-ray lines towards galaxy clusters with the
  Fermi-LAT}.
\newblock \emph{\jcap}, 2:\penalty0 026, February 2016.
\newblock \doi{10.1088/1475-7516/2016/02/026}.

\bibitem[{Quincy Adams} et~al.(2016){Quincy Adams}, {Bergstrom}, and
  {Spolyar}]{Quincy2016}
D.~{Quincy Adams}, L.~{Bergstrom}, and D.~{Spolyar}.
\newblock {Improved Constraints on Dark Matter Annihilation to a Line using
  Fermi-LAT observations of Galaxy Clusters}.
\newblock \emph{ArXiv e-prints}, June 2016.

\bibitem[{Bergstr{\"o}m} et~al.(2001){Bergstr{\"o}m}, {Edsj{\"o}}, and
  {Ullio}]{Bergstrom2001}
L.~{Bergstr{\"o}m}, J.~{Edsj{\"o}}, and P.~{Ullio}.
\newblock {Spectral Gamma-Ray Signatures of Cosmological Dark Matter
  Annihilations}.
\newblock \emph{Physical Review Letters}, 87\penalty0 (25):\penalty0 251301,
  December 2001.
\newblock \doi{10.1103/PhysRevLett.87.251301}.

\bibitem[{Ullio} et~al.(2002){Ullio}, {Bergstr{\"o}m}, {Edsj{\"o}}, and
  {Lacey}]{Ullio2002}
P.~{Ullio}, L.~{Bergstr{\"o}m}, J.~{Edsj{\"o}}, and C.~{Lacey}.
\newblock {Cosmological dark matter annihilations into {$\gamma$} rays: A
  closer look}.
\newblock \emph{\prd}, 66\penalty0 (12):\penalty0 123502, December 2002.
\newblock \doi{10.1103/PhysRevD.66.123502}.

\bibitem[{Taylor} and {Silk}(2003)]{Taylor2003}
J.~E. {Taylor} and J.~{Silk}.
\newblock {The clumpiness of cold dark matter: implications for the
  annihilation signal}.
\newblock \emph{\mnras}, 339:\penalty0 505--514, February 2003.
\newblock \doi{10.1046/j.1365-8711.2003.06201.x}.

\bibitem[{Di Mauro} and {Donato}(2015)]{DiMauro2015}
M.~{Di Mauro} and F.~{Donato}.
\newblock {Composition of the Fermi-LAT isotropic gamma-ray background
  intensity: Emission from extragalactic point sources and dark matter
  annihilations}.
\newblock \emph{\prd}, 91\penalty0 (12):\penalty0 123001, June 2015.
\newblock \doi{10.1103/PhysRevD.91.123001}.

\bibitem[{Ackermann, et al.}(2015{\natexlab{c}})]{Ackermann2015c}
M.~{Ackermann, et al.}
\newblock {The Spectrum of Isotropic Diffuse Gamma-Ray Emission between 100 MeV
  and 820 GeV}.
\newblock \emph{\apj}, 799:\penalty0 86, January 2015{\natexlab{c}}.
\newblock \doi{10.1088/0004-637X/799/1/86}.

\bibitem[{Ando} and {Pavlidou}(2009)]{Ando2009}
S.~{Ando} and V.~{Pavlidou}.
\newblock {Imprint of galaxy clustering in the cosmic gamma-ray background}.
\newblock \emph{\mnras}, 400:\penalty0 2122--2127, December 2009.
\newblock \doi{10.1111/j.1365-2966.2009.15605.x}.

\bibitem[{Camera} et~al.(2013){Camera}, {Fornasa}, {Fornengo}, and
  {Regis}]{Camera2013}
S.~{Camera}, M.~{Fornasa}, N.~{Fornengo}, and M.~{Regis}.
\newblock {A Novel Approach in the Weakly Interacting Massive Particle Quest:
  Cross-correlation of Gamma-Ray Anisotropies and Cosmic Shear}.
\newblock \emph{\apjl}, 771:\penalty0 L5, July 2013.
\newblock \doi{10.1088/2041-8205/771/1/L5}.

\bibitem[{Ando}(2014)]{Ando2014}
S.~{Ando}.
\newblock {Power spectrum tomography of dark matter annihilation with local
  galaxy distribution}.
\newblock \emph{\jcap}, 10:\penalty0 061, October 2014.
\newblock \doi{10.1088/1475-7516/2014/10/061}.

\bibitem[{Shirasaki} et~al.(2014){Shirasaki}, {Horiuchi}, and
  {Yoshida}]{Shirasaki2014}
M.~{Shirasaki}, S.~{Horiuchi}, and N.~{Yoshida}.
\newblock {Cross correlation of cosmic shear and extragalactic gamma-ray
  background: Constraints on the dark matter annihilation cross section}.
\newblock \emph{\prd}, 90\penalty0 (6):\penalty0 063502, September 2014.
\newblock \doi{10.1103/PhysRevD.90.063502}.

\bibitem[{Camera} et~al.(2015){Camera}, {Fornasa}, {Fornengo}, and
  {Regis}]{Camera2015}
S.~{Camera}, M.~{Fornasa}, N.~{Fornengo}, and M.~{Regis}.
\newblock {Tomographic-spectral approach for dark matter detection in the
  cross-correlation between cosmic shear and diffuse {$\gamma$}-ray emission}.
\newblock \emph{\jcap}, 6:\penalty0 029, June 2015.
\newblock \doi{10.1088/1475-7516/2015/06/029}.

\bibitem[{Cuoco} et~al.(2015){Cuoco}, {Xia}, {Regis}, {Branchini}, {Fornengo},
  and {Viel}]{Cuoco2015}
A.~{Cuoco}, J.-Q. {Xia}, M.~{Regis}, E.~{Branchini}, N.~{Fornengo}, and
  M.~{Viel}.
\newblock {Dark Matter Searches in the Gamma-ray Extragalactic Background via
  Cross-correlations with Galaxy Catalogs}.
\newblock \emph{\apjs}, 221:\penalty0 29, December 2015.
\newblock \doi{10.1088/0067-0049/221/2/29}.

\bibitem[{Abazajian} et~al.(2014){Abazajian}, {Canac}, {Horiuchi}, and
  {Kaplinghat}]{Abazajian2014}
K.~N. {Abazajian}, N.~{Canac}, S.~{Horiuchi}, and M.~{Kaplinghat}.
\newblock {Astrophysical and dark matter interpretations of extended gamma-ray
  emission from the Galactic Center}.
\newblock \emph{\prd}, 90\penalty0 (2):\penalty0 023526, July 2014.
\newblock \doi{10.1103/PhysRevD.90.023526}.

\bibitem[{van Eldik}(2015)]{vanEldik2015}
C.~{van Eldik}.
\newblock {Gamma rays from the Galactic Centre region: A review}.
\newblock \emph{Astroparticle Physics}, 71:\penalty0 45--70, December 2015.
\newblock \doi{10.1016/j.astropartphys.2015.05.002}.

\bibitem[{Goodenough} and {Hooper}(2009)]{Goodenough2009}
L.~{Goodenough} and D.~{Hooper}.
\newblock {Possible Evidence For Dark Matter Annihilation In The Inner Milky
  Way From The Fermi Gamma Ray Space Telescope}.
\newblock \emph{ArXiv e-prints}, October 2009.

\bibitem[{Hooper} and {Goodenough}(2011)]{Hooper2011}
D.~{Hooper} and L.~{Goodenough}.
\newblock {Dark matter annihilation in the Galactic Center as seen by the Fermi
  Gamma Ray Space Telescope}.
\newblock \emph{Physics Letters B}, 697:\penalty0 412--428, March 2011.
\newblock \doi{10.1016/j.physletb.2011.02.029}.

\bibitem[{Daylan, et al.}(2016)]{Daylan2016}
T.~{Daylan, et al.}
\newblock {The characterization of the gamma-ray signal from the central Milky
  Way: A case for annihilating dark matter}.
\newblock \emph{Physics of the Dark Universe}, 12:\penalty0 1--23, June 2016.
\newblock \doi{10.1016/j.dark.2015.12.005}.

\bibitem[{Baltz} et~al.(2007){Baltz}, {Taylor}, and {Wai}]{Baltz2007}
E.~A. {Baltz}, J.~E. {Taylor}, and L.~L. {Wai}.
\newblock {Can Astrophysical Gamma-Ray Sources Mimic Dark Matter Annihilation
  in Galactic Satellites?}
\newblock \emph{\apjl}, 659:\penalty0 L125--L128, April 2007.
\newblock \doi{10.1086/517882}.

\bibitem[{Sawala, et al.}(2016)]{Sawala2016}
T.~{Sawala, et al.}
\newblock {The APOSTLE simulations: solutions to the Local Group's cosmic
  puzzles}.
\newblock \emph{\mnras}, 457:\penalty0 1931--1943, April 2016.
\newblock \doi{10.1093/mnras/stw145}.

\bibitem[{Oman, et al.}(2015)]{Oman2015}
K.~A. {Oman, et al.}
\newblock {The unexpected diversity of dwarf galaxy rotation curves}.
\newblock \emph{\mnras}, 452:\penalty0 3650--3665, October 2015.
\newblock \doi{10.1093/mnras/stv1504}.

\bibitem[{S{\'a}nchez-Conde} et~al.(2011){S{\'a}nchez-Conde}, {Cannoni},
  {Zandanel}, {G{\'o}mez}, and {Prada}]{Sanchez-Conde2011}
M.~A. {S{\'a}nchez-Conde}, M.~{Cannoni}, F.~{Zandanel}, M.~E. {G{\'o}mez}, and
  F.~{Prada}.
\newblock {Dark matter searches with Cherenkov telescopes: nearby dwarf
  galaxies or local galaxy clusters?}
\newblock \emph{\jcap}, 12:\penalty0 011, December 2011.
\newblock \doi{10.1088/1475-7516/2011/12/011}.

\bibitem[{Lisanti} et~al.(2017){Lisanti}, {Mishra-Sharma}, {Rodd}, and
  {Safdi}]{Lisanti2017}
M.~{Lisanti}, S.~{Mishra-Sharma}, N.~L. {Rodd}, and B.~R. {Safdi}.
\newblock {A Search for Dark Matter Annihilation in Galaxy Groups}.
\newblock \emph{ArXiv e-prints}, August 2017.

\bibitem[{Ackermann, et al.}(2017)]{Ackermann2017}
M.~{Ackermann, et al.}
\newblock {Observations of M31 and M33 with the Fermi Large Area Telescope: A
  Galactic Center Excess in Andromeda?}
\newblock \emph{\apj}, 836:\penalty0 208, February 2017.
\newblock \doi{10.3847/1538-4357/aa5c3d}.

\bibitem[{Lavalle} et~al.(2008){Lavalle}, {Yuan}, {Maurin}, and
  {Bi}]{Lavalle2008}
J.~{Lavalle}, Q.~{Yuan}, D.~{Maurin}, and X.-J. {Bi}.
\newblock {Full calculation of clumpiness boost factors for antimatter cosmic
  rays in the light of {$\Lambda$}CDM N-body simulation results. Abandoning
  hope in clumpiness enhancement?}
\newblock \emph{\aap}, 479:\penalty0 427--452, February 2008.
\newblock \doi{10.1051/0004-6361:20078723}.

\bibitem[{Cirelli, et al.}(2011)]{Cirelli2011}
M.~{Cirelli, et al.}
\newblock {PPPC 4 DM ID: a poor particle physicist cookbook for dark matter
  indirect detection}.
\newblock \emph{\jcap}, 3:\penalty0 051, March 2011.
\newblock \doi{10.1088/1475-7516/2011/03/051}.

\bibitem[{Navarro} et~al.(1996){Navarro}, {Frenk}, and {White}]{NFW96}
J.~F. {Navarro}, C.~S. {Frenk}, and S.~D.~M. {White}.
\newblock {The Structure of Cold Dark Matter Halos}.
\newblock \emph{apj}, 462:\penalty0 563, May 1996.
\newblock \doi{10.1086/177173}.

\bibitem[{Navarro} et~al.(2004){Navarro}, {Hayashi}, {Power}, {Jenkins},
  {Frenk}, {White}, {Springel}, {Stadel}, and {Quinn}]{Navarro2004}
J.~F. {Navarro}, E.~{Hayashi}, C.~{Power}, A.~R. {Jenkins}, C.~S. {Frenk},
  S.~D.~M. {White}, V.~{Springel}, J.~{Stadel}, and T.~R. {Quinn}.
\newblock {The inner structure of {$\Lambda$}CDM haloes - III. Universality and
  asymptotic slopes}.
\newblock \emph{\mnras}, 349:\penalty0 1039--1051, April 2004.
\newblock \doi{10.1111/j.1365-2966.2004.07586.x}.

\bibitem[{Merritt} et~al.(2006){Merritt}, {Graham}, {Moore}, {Diemand}, and
  {Terzi{\'c}}]{Merritt2006}
D.~{Merritt}, A.~W. {Graham}, B.~{Moore}, J.~{Diemand}, and B.~{Terzi{\'c}}.
\newblock {Empirical Models for Dark Matter Halos. I. Nonparametric
  Construction of Density Profiles and Comparison with Parametric Models}.
\newblock \emph{\aj}, 132:\penalty0 2685--2700, December 2006.
\newblock \doi{10.1086/508988}.

\bibitem[{Gao} et~al.(2008){Gao}, {Navarro}, {Cole}, {Frenk}, {White},
  {Springel}, {Jenkins}, and {Neto}]{Gao2008}
L.~{Gao}, J.~F. {Navarro}, S.~{Cole}, C.~S. {Frenk}, S.~D.~M. {White},
  V.~{Springel}, A.~{Jenkins}, and A.~F. {Neto}.
\newblock {The redshift dependence of the structure of massive {$\Lambda$} cold
  dark matter haloes}.
\newblock \emph{\mnras}, 387:\penalty0 536--544, June 2008.
\newblock \doi{10.1111/j.1365-2966.2008.13277.x}.

\bibitem[{Klypin} et~al.(2016){Klypin}, {Yepes}, {Gottl{\"o}ber}, {Prada}, and
  {He{\ss}}]{Klypin2016}
A.~{Klypin}, G.~{Yepes}, S.~{Gottl{\"o}ber}, F.~{Prada}, and S.~{He{\ss}}.
\newblock {MultiDark simulations: the story of dark matter halo concentrations
  and density profiles}.
\newblock \emph{\mnras}, 457:\penalty0 4340--4359, April 2016.
\newblock \doi{10.1093/mnras/stw248}.

\bibitem[{Prada} et~al.(2012){Prada}, {Klypin}, {Cuesta}, {Betancort-Rijo}, and
  {Primack}]{Prada2012}
F.~{Prada}, A.~A. {Klypin}, A.~J. {Cuesta}, J.~E. {Betancort-Rijo}, and
  J.~{Primack}.
\newblock {Halo concentrations in the standard {$\Lambda$} cold dark matter
  cosmology}.
\newblock \emph{mnras}, 423:\penalty0 3018--3030, July 2012.
\newblock \doi{10.1111/j.1365-2966.2012.21007.x}.

\bibitem[{Alam} et~al.(2002){Alam}, {Bullock}, and {Weinberg}]{Alam2002}
S.~M.~K. {Alam}, J.~S. {Bullock}, and D.~H. {Weinberg}.
\newblock {Dark Matter Properties and Halo Central Densities}.
\newblock \emph{\apj}, 572:\penalty0 34--40, June 2002.
\newblock \doi{10.1086/340190}.

\bibitem[{Diemand} et~al.(2007){Diemand}, {Kuhlen}, and {Madau}]{Diemand2007}
J.~{Diemand}, M.~{Kuhlen}, and P.~{Madau}.
\newblock {Formation and Evolution of Galaxy Dark Matter Halos and Their
  Substructure}.
\newblock \emph{\apj}, 667:\penalty0 859--877, October 2007.
\newblock \doi{10.1086/520573}.

\bibitem[{Molin{\'e}} et~al.(2017){Molin{\'e}}, {S{\'a}nchez-Conde},
  {Palomares-Ruiz}, and {Prada}]{Moline2017}
{\'A}.~{Molin{\'e}}, M.~A. {S{\'a}nchez-Conde}, S.~{Palomares-Ruiz}, and
  F.~{Prada}.
\newblock {Characterization of subhalo structural properties and implications
  for dark matter annihilation signals}.
\newblock \emph{\mnras}, 466:\penalty0 4974--4990, April 2017.
\newblock \doi{10.1093/mnras/stx026}.

\bibitem[{Bullock, et al.}(2001)]{Bullock2001}
J.~S. {Bullock, et al.}
\newblock {Profiles of dark haloes: evolution, scatter and environment}.
\newblock \emph{mnras}, 321:\penalty0 559--575, March 2001.
\newblock \doi{10.1046/j.1365-8711.2001.04068.x}.

\bibitem[{Macci{\`o}} et~al.(2008){Macci{\`o}}, {Dutton}, and {van den
  Bosch}]{Maccio2008}
A.~V. {Macci{\`o}}, A.~A. {Dutton}, and F.~C. {van den Bosch}.
\newblock {Concentration, spin and shape of dark matter haloes as a function of
  the cosmological model: WMAP1, WMAP3 and WMAP5 results}.
\newblock \emph{\mnras}, 391:\penalty0 1940--1954, December 2008.
\newblock \doi{10.1111/j.1365-2966.2008.14029.x}.

\bibitem[Zhao et~al.(2009)Zhao, Jing, Mo, and Börner]{Zhao09}
D.~H. Zhao, Y.~P. Jing, H.~J. Mo, and G.~Börner.
\newblock Accurate universal models for the mass accretion histories and
  concentrations of dark matter halos.
\newblock \emph{The Astrophysical Journal}, 707\penalty0 (1):\penalty0 354,
  2009.

\bibitem[{Klypin} et~al.(2011){Klypin}, {Trujillo-Gomez}, and
  {Primack}]{Klypin2011}
A.~A. {Klypin}, S.~{Trujillo-Gomez}, and J.~{Primack}.
\newblock {Dark Matter Halos in the Standard Cosmological Model: Results from
  the Bolshoi Simulation}.
\newblock \emph{apj}, 740:\penalty0 102, October 2011.
\newblock \doi{10.1088/0004-637X/740/2/102}.

\bibitem[{Dutton} and {Macci{\`o}}(2014)]{Dutton2014}
A.~A. {Dutton} and A.~V. {Macci{\`o}}.
\newblock {Cold dark matter haloes in the Planck era: evolution of structural
  parameters for Einasto and NFW profiles}.
\newblock \emph{mnras}, 441:\penalty0 3359--3374, July 2014.
\newblock \doi{10.1093/mnras/stu742}.

\bibitem[{Diemer} and {Kravtsov}(2015)]{Diemer2015}
B.~{Diemer} and A.~V. {Kravtsov}.
\newblock {A Universal Model for Halo Concentrations}.
\newblock \emph{apj}, 799:\penalty0 108, January 2015.
\newblock \doi{10.1088/0004-637X/799/1/108}.

\bibitem[{Correa} et~al.(2015){Correa}, {Wyithe}, {Schaye}, and
  {Duffy}]{Correa2015}
C.~A. {Correa}, J.~S.~B. {Wyithe}, J.~{Schaye}, and A.~R. {Duffy}.
\newblock {The accretion history of dark matter haloes - III. A physical model
  for the concentration-mass relation}.
\newblock \emph{\mnras}, 452:\penalty0 1217--1232, September 2015.
\newblock \doi{10.1093/mnras/stv1363}.

\bibitem[{Ludlow} et~al.(2012){Ludlow}, {Navarro}, {Li}, {Angulo},
  {Boylan-Kolchin}, and {Bett}]{Ludlow2012}
A.~D. {Ludlow}, J.~F. {Navarro}, M.~{Li}, R.~E. {Angulo}, M.~{Boylan-Kolchin},
  and P.~E. {Bett}.
\newblock {The dynamical state and mass-concentration relation of galaxy
  clusters}.
\newblock \emph{mnras}, 427:\penalty0 1322--1328, December 2012.
\newblock \doi{10.1111/j.1365-2966.2012.21892.x}.

\bibitem[{Pinzke} et~al.(2011){Pinzke}, {Pfrommer}, and
  {Bergstr{\"o}m}]{Pinzke2011}
A.~{Pinzke}, C.~{Pfrommer}, and L.~{Bergstr{\"o}m}.
\newblock {Prospects of detecting gamma-ray emission from galaxy clusters:
  Cosmic rays and dark matter annihilations}.
\newblock \emph{\prd}, 84\penalty0 (12):\penalty0 123509, December 2011.
\newblock \doi{10.1103/PhysRevD.84.123509}.

\bibitem[{Gao} et~al.(2012){Gao}, {Frenk}, {Jenkins}, {Springel}, and
  {White}]{Gao2012}
L.~{Gao}, C.~S. {Frenk}, A.~{Jenkins}, V.~{Springel}, and S.~D.~M. {White}.
\newblock {Where will supersymmetric dark matter first be seen?}
\newblock \emph{\mnras}, 419:\penalty0 1721--1726, January 2012.
\newblock \doi{10.1111/j.1365-2966.2011.19836.x}.

\bibitem[{S{\'a}nchez-Conde} and {Prada}(2014)]{Sanchez-Conde2014}
M.~A. {S{\'a}nchez-Conde} and F.~{Prada}.
\newblock {The flattening of the concentration-mass relation towards low halo
  masses and its implications for the annihilation signal boost}.
\newblock \emph{\mnras}, 442:\penalty0 2271--2277, August 2014.
\newblock \doi{10.1093/mnras/stu1014}.

\bibitem[{Okoli} and {Afshordi}(2016)]{Okoli2016}
C.~{Okoli} and N.~{Afshordi}.
\newblock {Concentration, ellipsoidal collapse, and the densest dark matter
  haloes}.
\newblock \emph{mnras}, 456:\penalty0 3068--3078, March 2016.
\newblock \doi{10.1093/mnras/stv2905}.

\bibitem[{Ishiyama}(2014)]{Ishiyama2014}
T.~{Ishiyama}.
\newblock {Hierarchical Formation of Dark Matter Halos and the Free Streaming
  Scale}.
\newblock \emph{apj}, 788:\penalty0 27, June 2014.
\newblock \doi{10.1088/0004-637X/788/1/27}.

\bibitem[{Pilipenko} et~al.(2017){Pilipenko}, {S{\'a}nchez-Conde}, {Prada}, and
  {Yepes}]{Pilipenko2017}
S.~V. {Pilipenko}, M.~A. {S{\'a}nchez-Conde}, F.~{Prada}, and G.~{Yepes}.
\newblock {Pushing down the low-mass halo concentration frontier with the
  Lomonosov cosmological simulations}.
\newblock \emph{\mnras}, 472:\penalty0 4918--4927, August 2017.
\newblock \doi{10.1093/mnras/stx2319}.

\bibitem[{Bringmann}(2009)]{Bringmann2009}
T.~{Bringmann}.
\newblock {Particle models and the small-scale structure of dark matter}.
\newblock \emph{New Journal of Physics}, 11\penalty0 (10):\penalty0 105027,
  October 2009.
\newblock \doi{10.1088/1367-2630/11/10/105027}.

\bibitem[{Diemand} et~al.(2005){Diemand}, {Moore}, and {Stadel}]{Diemand2005}
J.~{Diemand}, B.~{Moore}, and J.~{Stadel}.
\newblock {Earth-mass dark-matter haloes as the first structures in the early
  Universe}.
\newblock \emph{\nat}, 433:\penalty0 389--391, January 2005.
\newblock \doi{10.1038/nature03270}.

\bibitem[{Ishiyama} et~al.(2010){Ishiyama}, {Makino}, and
  {Ebisuzaki}]{Ishiyama2010}
T.~{Ishiyama}, J.~{Makino}, and T.~{Ebisuzaki}.
\newblock {Gamma-ray Signal from Earth-mass Dark Matter Microhalos}.
\newblock \emph{\apjl}, 723:\penalty0 L195--L200, November 2010.
\newblock \doi{10.1088/2041-8205/723/2/L195}.

\bibitem[{Anderhalden} and {Diemand}(2013)]{Anderhalden2013}
D.~{Anderhalden} and J.~{Diemand}.
\newblock {Density profiles of CDM microhalos and their implications for
  annihilation boost factors}.
\newblock \emph{jcap}, 4:\penalty0 009, April 2013.
\newblock \doi{10.1088/1475-7516/2013/04/009}.

\bibitem[{Moore} et~al.(2004){Moore}, {Kazantzidis}, {Diemand}, and
  {Stadel}]{Moore2004}
B.~{Moore}, S.~{Kazantzidis}, J.~{Diemand}, and J.~{Stadel}.
\newblock {The origin and tidal evolution of cuspy triaxial haloes}.
\newblock \emph{\mnras}, 354:\penalty0 522--528, October 2004.
\newblock \doi{10.1111/j.1365-2966.2004.08211.x}.

\bibitem[{Han} et~al.(2016){Han}, {Cole}, {Frenk}, and {Jing}]{Han2016}
J.~{Han}, S.~{Cole}, C.~S. {Frenk}, and Y.~{Jing}.
\newblock {A unified model for the spatial and mass distribution of subhaloes}.
\newblock \emph{\mnras}, 457:\penalty0 1208--1223, April 2016.
\newblock \doi{10.1093/mnras/stv2900}.

\bibitem[{Bartels} and {Ando}(2015)]{Bartels2015}
R.~{Bartels} and S.~{Ando}.
\newblock {Boosting the annihilation boost: Tidal effects on dark matter
  subhalos and consistent luminosity modeling}.
\newblock \emph{\prd}, 92\penalty0 (12):\penalty0 123508, December 2015.
\newblock \doi{10.1103/PhysRevD.92.123508}.

\bibitem[{Stref} and {Lavalle}(2017)]{Stref2017}
M.~{Stref} and J.~{Lavalle}.
\newblock {Modeling dark matter subhalos in a constrained galaxy: Global mass
  and boosted annihilation profiles}.
\newblock \emph{\prd}, 95\penalty0 (6):\penalty0 063003, March 2017.
\newblock \doi{10.1103/PhysRevD.95.063003}.

\bibitem[{H{\"u}tten} et~al.(2018){H{\"u}tten}, {Combet}, and
  {Maurin}]{Hutten2018}
M.~{H{\"u}tten}, C.~{Combet}, and D.~{Maurin}.
\newblock {Extragalactic diffuse {$\gamma$}-rays from dark matter annihilation:
  revised prediction and full modelling uncertainties}.
\newblock \emph{\jcap}, 2:\penalty0 005, February 2018.
\newblock \doi{10.1088/1475-7516/2018/02/005}.

\bibitem[{Kamionkowski} and {Koushiappas}(2008)]{Kamionkowski2008}
M.~{Kamionkowski} and S.~M. {Koushiappas}.
\newblock {Galactic substructure and direct detection of dark matter}.
\newblock \emph{\prd}, 77\penalty0 (10):\penalty0 103509, May 2008.
\newblock \doi{10.1103/PhysRevD.77.103509}.

\bibitem[{Kamionkowski} et~al.(2010){Kamionkowski}, {Koushiappas}, and
  {Kuhlen}]{Kamionkowski2010}
M.~{Kamionkowski}, S.~M. {Koushiappas}, and M.~{Kuhlen}.
\newblock {Galactic substructure and dark-matter annihilation in the Milky Way
  halo}.
\newblock \emph{\prd}, 81\penalty0 (4):\penalty0 043532, February 2010.
\newblock \doi{10.1103/PhysRevD.81.043532}.

\bibitem[{Zavala} and {Afshordi}(2014)]{P2SAD}
J.~{Zavala} and N.~{Afshordi}.
\newblock {Clustering in the phase space of dark matter haloes - II. Stable
  clustering and dark matter annihilation}.
\newblock \emph{mnras}, 441:\penalty0 1329--1339, June 2014.
\newblock \doi{10.1093/mnras/stu506}.

\bibitem[{Gao} et~al.(2011){Gao}, {Frenk}, {Boylan-Kolchin}, {Jenkins},
  {Springel}, and {White}]{2011MNRAS.410.2309G}
L.~{Gao}, C.~S. {Frenk}, M.~{Boylan-Kolchin}, A.~{Jenkins}, V.~{Springel}, and
  S.~D.~M. {White}.
\newblock {The statistics of the subhalo abundance of dark matter haloes}.
\newblock \emph{\mnras}, 410:\penalty0 2309--2314, February 2011.
\newblock \doi{10.1111/j.1365-2966.2010.17601.x}.

\bibitem[{Jiang} and {van den Bosch}(2014)]{Jiang2014}
F.~{Jiang} and F.~C. {van den Bosch}.
\newblock {Statistics of Dark Matter Substructure: I. Model and Universal
  Fitting Functions}.
\newblock \emph{ArXiv e-prints}, March 2014.

\bibitem[{Taylor} and {Babul}(2005)]{Taylor2005}
J.~E. {Taylor} and A.~{Babul}.
\newblock {The evolution of substructure in galaxy, group and cluster haloes -
  II. Global properties}.
\newblock \emph{\mnras}, 364:\penalty0 515--534, December 2005.
\newblock \doi{10.1111/j.1365-2966.2005.09582.x}.

\bibitem[{van den Bosch} et~al.(2018){van den Bosch}, {Ogiya}, {Hahn}, and
  {Burkert}]{vdB2018}
F.~C. {van den Bosch}, G.~{Ogiya}, O.~{Hahn}, and A.~{Burkert}.
\newblock {Disruption of dark matter substructure: fact or fiction?}
\newblock \emph{\mnras}, 474:\penalty0 3043--3066, March 2018.
\newblock \doi{10.1093/mnras/stx2956}.

\bibitem[{Hayashi} et~al.(2003){Hayashi}, {Navarro}, {Taylor}, {Stadel}, and
  {Quinn}]{Hayashi2003}
E.~{Hayashi}, J.~F. {Navarro}, J.~E. {Taylor}, J.~{Stadel}, and T.~{Quinn}.
\newblock {The Structural Evolution of Substructure}.
\newblock \emph{apj}, 584:\penalty0 541--558, February 2003.
\newblock \doi{10.1086/345788}.

\bibitem[{Drakos} et~al.(2017){Drakos}, {Taylor}, and {Benson}]{Drakos2017}
N.~E. {Drakos}, J.~E. {Taylor}, and A.~J. {Benson}.
\newblock {The phase-space structure of tidally stripped haloes}.
\newblock \emph{\mnras}, 468:\penalty0 2345--2358, June 2017.
\newblock \doi{10.1093/mnras/stx652}.

\bibitem[Wechsler et~al.(2002)Wechsler, Bullock, Primack, Kravtsov, and
  Dekel]{Wechsler2002}
Risa~H. Wechsler, James~S. Bullock, Joel~R. Primack, Andrey~V. Kravtsov, and
  Avishai Dekel.
\newblock Concentrations of dark halos from their assembly histories.
\newblock \emph{The Astrophysical Journal}, 568\penalty0 (1):\penalty0 52,
  2002.

\bibitem[{Reed} et~al.(2007){Reed}, {Bower}, {Frenk}, {Jenkins}, and
  {Theuns}]{Reed2007}
D.~S. {Reed}, R.~{Bower}, C.~S. {Frenk}, A.~{Jenkins}, and T.~{Theuns}.
\newblock {The halo mass function from the dark ages through the present day}.
\newblock \emph{mnras}, 374:\penalty0 2--15, January 2007.
\newblock \doi{10.1111/j.1365-2966.2006.11204.x}.

\bibitem[{Hofmann} et~al.(2001){Hofmann}, {Schwarz}, and
  {St{\"o}cker}]{Hofmann2001}
S.~{Hofmann}, D.~J. {Schwarz}, and H.~{St{\"o}cker}.
\newblock {Damping scales of neutralino cold dark matter}.
\newblock \emph{\prd}, 64\penalty0 (8):\penalty0 083507, October 2001.
\newblock \doi{10.1103/PhysRevD.64.083507}.

\bibitem[{Berezinsky} et~al.(2003){Berezinsky}, {Dokuchaev}, and
  {Eroshenko}]{Berezinsky2003}
V.~{Berezinsky}, V.~{Dokuchaev}, and Y.~{Eroshenko}.
\newblock {Small-scale clumps in the galactic halo and dark matter
  annihilation}.
\newblock \emph{\prd}, 68\penalty0 (10):\penalty0 103003, November 2003.
\newblock \doi{10.1103/PhysRevD.68.103003}.

\bibitem[{Green} et~al.(2004){Green}, {Hofmann}, and {Schwarz}]{Green2004}
A.~M. {Green}, S.~{Hofmann}, and D.~J. {Schwarz}.
\newblock {The power spectrum of SUSY-CDM on subgalactic scales}.
\newblock \emph{\mnras}, 353:\penalty0 L23--L27, September 2004.
\newblock \doi{10.1111/j.1365-2966.2004.08232.x}.

\bibitem[{Loeb} and {Zaldarriaga}(2005)]{Loeb2005}
A.~{Loeb} and M.~{Zaldarriaga}.
\newblock {Small-scale power spectrum of cold dark matter}.
\newblock \emph{\prd}, 71\penalty0 (10):\penalty0 103520, May 2005.
\newblock \doi{10.1103/PhysRevD.71.103520}.

\bibitem[{Bertschinger}(2006)]{Bertschinger2006}
E.~{Bertschinger}.
\newblock {Effects of cold dark matter decoupling and pair annihilation on
  cosmological perturbations}.
\newblock \emph{\prd}, 74\penalty0 (6):\penalty0 063509, September 2006.
\newblock \doi{10.1103/PhysRevD.74.063509}.

\bibitem[{Profumo} et~al.(2006){Profumo}, {Sigurdson}, and
  {Kamionkowski}]{Profumo2006}
S.~{Profumo}, K.~{Sigurdson}, and M.~{Kamionkowski}.
\newblock {What Mass Are the Smallest Protohalos?}
\newblock \emph{Physical Review Letters}, 97\penalty0 (3):\penalty0 031301,
  July 2006.
\newblock \doi{10.1103/PhysRevLett.97.031301}.

\bibitem[{Bringmann} and {Hofmann}(2007)]{Bringmann2007}
T.~{Bringmann} and S.~{Hofmann}.
\newblock {Thermal decoupling of WIMPs from first principles}.
\newblock \emph{\jcap}, 4:\penalty0 016, April 2007.
\newblock \doi{10.1088/1475-7516/2007/04/016}.

\bibitem[{Yang} et~al.(2011){Yang}, {Mo}, {Zhang}, and {van den
  Bosch}]{Yang2011}
X.~{Yang}, H.~J. {Mo}, Y.~{Zhang}, and F.~C. {van den Bosch}.
\newblock {An Analytical Model for the Accretion of Dark Matter Subhalos}.
\newblock \emph{\apj}, 741:\penalty0 13, November 2011.
\newblock \doi{10.1088/0004-637X/741/1/13}.

\bibitem[{Ludlow} et~al.(2014){Ludlow}, {Navarro}, {Angulo}, {Boylan-Kolchin},
  {Springel}, {Frenk}, and {White}]{2014MNRAS.441..378L}
A.~D. {Ludlow}, J.~F. {Navarro}, R.~E. {Angulo}, M.~{Boylan-Kolchin},
  V.~{Springel}, C.~{Frenk}, and S.~D.~M. {White}.
\newblock {The mass-concentration-redshift relation of cold dark matter
  haloes}.
\newblock \emph{mnras}, 441:\penalty0 378--388, June 2014.
\newblock \doi{10.1093/mnras/stu483}.

\bibitem[{Abdo, et al.}(2010)]{Abdo2010}
A.~A. {Abdo, et al.}
\newblock {Fermi Large Area Telescope observations of Local Group galaxies:
  detection of M 31 and search for M 33}.
\newblock \emph{aap}, 523:\penalty0 L2, November 2010.
\newblock \doi{10.1051/0004-6361/201015759}.

\bibitem[{Selig} et~al.(2015){Selig}, {Vacca}, {Oppermann}, and
  {En{\ss}lin}]{Selig2015}
M.~{Selig}, V.~{Vacca}, N.~{Oppermann}, and T.~A. {En{\ss}lin}.
\newblock {The denoised, deconvolved, and decomposed Fermi {$\gamma$}-ray sky.
  An application of the D$^{3}$PO algorithm}.
\newblock \emph{\aap}, 581:\penalty0 A126, September 2015.
\newblock \doi{10.1051/0004-6361/201425172}.

\bibitem[{Ackermann} et~al.(2012){Ackermann}, {Ajello}, {Allafort}, {Baldini},
  {Ballet}, {Bastieri}, {Bechtol}, {Bellazzini}, {Berenji}, {Bloom},
  {Bonamente}, {Borgland}, {Bouvier}, {Bregeon}, {Brigida}, {Bruel}, {Buehler},
  {Buson}, {Caliandro}, {Cameron}, {Caraveo}, {Casandjian}, {Cecchi},
  {Charles}, {Chekhtman}, {Cheung}, {Chiang}, {Cillis}, {Ciprini}, {Claus},
  {Cohen-Tanugi}, {Conrad}, {Cutini}, {de Palma}, {Dermer}, {Digel}, {Silva},
  {Drell}, {Drlica-Wagner}, {Favuzzi}, {Fegan}, {Fortin}, {Fukazawa}, {Funk},
  {Fusco}, {Gargano}, {Gasparrini}, {Germani}, {Giglietto}, {Giordano},
  {Glanzman}, {Godfrey}, {Grenier}, {Guiriec}, {Gustafsson}, {Hadasch},
  {Hayashida}, {Hays}, {Hughes}, {J{\'o}hannesson}, {Johnson}, {Kamae},
  {Katagiri}, {Kataoka}, {Kn{\"o}dlseder}, {Kuss}, {Lande}, {Longo}, {Loparco},
  {Lott}, {Lovellette}, {Lubrano}, {Madejski}, {Martin}, {Mazziotta},
  {McEnery}, {Michelson}, {Mizuno}, {Monte}, {Monzani}, {Morselli},
  {Moskalenko}, {Murgia}, {Nishino}, {Norris}, {Nuss}, {Ohno}, {Ohsugi},
  {Okumura}, {Omodei}, {Orlando}, {Ozaki}, {Parent}, {Persic}, {Pesce-Rollins},
  {Petrosian}, {Pierbattista}, {Piron}, {Pivato}, {Porter}, {Rain{\`o}},
  {Rando}, {Razzano}, {Reimer}, {Reimer}, {Ritz}, {Roth}, {Sbarra}, {Sgr{\`o}},
  {Siskind}, {Spandre}, {Spinelli}, {Stawarz}, {Strong}, {Takahashi}, {Tanaka},
  {Thayer}, {Tibaldo}, {Tinivella}, {Torres}, {Tosti}, {Troja}, {Uchiyama},
  {Vandenbroucke}, {Vianello}, {Vitale}, {Waite}, {Wood}, and
  {Yang}]{Ackermann2012}
M.~{Ackermann}, M.~{Ajello}, A.~{Allafort}, L.~{Baldini}, J.~{Ballet},
  D.~{Bastieri}, K.~{Bechtol}, R.~{Bellazzini}, B.~{Berenji}, E.~D. {Bloom},
  E.~{Bonamente}, A.~W. {Borgland}, A.~{Bouvier}, J.~{Bregeon}, M.~{Brigida},
  P.~{Bruel}, R.~{Buehler}, S.~{Buson}, G.~A. {Caliandro}, R.~A. {Cameron},
  P.~A. {Caraveo}, J.~M. {Casandjian}, C.~{Cecchi}, E.~{Charles},
  A.~{Chekhtman}, C.~C. {Cheung}, J.~{Chiang}, A.~N. {Cillis}, S.~{Ciprini},
  R.~{Claus}, J.~{Cohen-Tanugi}, J.~{Conrad}, S.~{Cutini}, F.~{de Palma}, C.~D.
  {Dermer}, S.~W. {Digel}, E.~d.~C.~e. {Silva}, P.~S. {Drell},
  A.~{Drlica-Wagner}, C.~{Favuzzi}, S.~J. {Fegan}, P.~{Fortin}, Y.~{Fukazawa},
  S.~{Funk}, P.~{Fusco}, F.~{Gargano}, D.~{Gasparrini}, S.~{Germani},
  N.~{Giglietto}, F.~{Giordano}, T.~{Glanzman}, G.~{Godfrey}, I.~A. {Grenier},
  S.~{Guiriec}, M.~{Gustafsson}, D.~{Hadasch}, M.~{Hayashida}, E.~{Hays}, R.~E.
  {Hughes}, G.~{J{\'o}hannesson}, A.~S. {Johnson}, T.~{Kamae}, H.~{Katagiri},
  J.~{Kataoka}, J.~{Kn{\"o}dlseder}, M.~{Kuss}, J.~{Lande}, F.~{Longo},
  F.~{Loparco}, B.~{Lott}, M.~N. {Lovellette}, P.~{Lubrano}, G.~M. {Madejski},
  P.~{Martin}, M.~N. {Mazziotta}, J.~E. {McEnery}, P.~F. {Michelson},
  T.~{Mizuno}, C.~{Monte}, M.~E. {Monzani}, A.~{Morselli}, I.~V. {Moskalenko},
  S.~{Murgia}, S.~{Nishino}, J.~P. {Norris}, E.~{Nuss}, M.~{Ohno}, T.~{Ohsugi},
  A.~{Okumura}, N.~{Omodei}, E.~{Orlando}, M.~{Ozaki}, D.~{Parent},
  M.~{Persic}, M.~{Pesce-Rollins}, V.~{Petrosian}, M.~{Pierbattista},
  F.~{Piron}, G.~{Pivato}, T.~A. {Porter}, S.~{Rain{\`o}}, R.~{Rando},
  M.~{Razzano}, A.~{Reimer}, O.~{Reimer}, S.~{Ritz}, M.~{Roth}, C.~{Sbarra},
  C.~{Sgr{\`o}}, E.~J. {Siskind}, G.~{Spandre}, P.~{Spinelli}, {\L}.~{Stawarz},
  A.~W. {Strong}, H.~{Takahashi}, T.~{Tanaka}, J.~B. {Thayer}, L.~{Tibaldo},
  M.~{Tinivella}, D.~F. {Torres}, G.~{Tosti}, E.~{Troja}, Y.~{Uchiyama},
  J.~{Vandenbroucke}, G.~{Vianello}, V.~{Vitale}, A.~P. {Waite}, M.~{Wood}, and
  Z.~{Yang}.
\newblock {GeV Observations of Star-forming Galaxies with the Fermi Large Area
  Telescope}.
\newblock \emph{\apj}, 755:\penalty0 164, August 2012.
\newblock \doi{10.1088/0004-637X/755/2/164}.

\bibitem[{Tamborra} et~al.(2014){Tamborra}, {Ando}, and {Murase}]{Tamborra2014}
I.~{Tamborra}, S.~{Ando}, and K.~{Murase}.
\newblock {Star-forming galaxies as the origin of diffuse high-energy
  backgrounds: gamma-ray and neutrino connections, and implications for
  starburst history}.
\newblock \emph{\jcap}, 9:\penalty0 043, September 2014.
\newblock \doi{10.1088/1475-7516/2014/09/043}.

\bibitem[{Bechtol} et~al.(2017){Bechtol}, {Ahlers}, {Di Mauro}, {Ajello}, and
  {Vandenbroucke}]{Bechtol2017}
K.~{Bechtol}, M.~{Ahlers}, M.~{Di Mauro}, M.~{Ajello}, and J.~{Vandenbroucke}.
\newblock {Evidence against Star-forming Galaxies as the Dominant Source of
  Icecube Neutrinos}.
\newblock \emph{\apj}, 836:\penalty0 47, February 2017.
\newblock \doi{10.3847/1538-4357/836/1/47}.

\bibitem[{Pfrommer} et~al.(2017){Pfrommer}, {Pakmor}, {Simpson}, and
  {Springel}]{Pfrommer2017}
C.~{Pfrommer}, R.~{Pakmor}, C.~M. {Simpson}, and V.~{Springel}.
\newblock {Simulating Gamma-Ray Emission in Star-forming Galaxies}.
\newblock \emph{\apjl}, 847:\penalty0 L13, October 2017.
\newblock \doi{10.3847/2041-8213/aa8bb1}.

\bibitem[{Behroozi} et~al.(2013){Behroozi}, {Wechsler}, and
  {Conroy}]{Behroozi2013}
P.~S. {Behroozi}, R.~H. {Wechsler}, and C.~{Conroy}.
\newblock {The Average Star Formation Histories of Galaxies in Dark Matter
  Halos from z = 0-8}.
\newblock \emph{apj}, 770:\penalty0 57, June 2013.
\newblock \doi{10.1088/0004-637X/770/1/57}.

\bibitem[{Santini, et al.}(2017)]{Santini2017}
P.~{Santini, et al.}
\newblock {The Main Sequence relation in the HST Frontier Fields}.
\newblock \emph{ArXiv e-prints}, June 2017.

\bibitem[{Acharya, et al.}(2013)]{Acharya2013}
B.~S. {Acharya, et al.}
\newblock {Introducing the CTA concept}.
\newblock \emph{Astroparticle Physics}, 43:\penalty0 3--18, March 2013.
\newblock \doi{10.1016/j.astropartphys.2013.01.007}.

\end{thebibliography}
\end{document}